\documentclass[journal]{IEEEtran}  

\IEEEoverridecommandlockouts                              

\usepackage{mathtools}    
\usepackage{amsfonts}
\usepackage{graphicx}   
\usepackage{subcaption} 
\usepackage{epsfig} 
\usepackage{cancel}
\usepackage{amssymb}
\usepackage{color}
\usepackage{bm}
\usepackage[ruled,vlined,titlenotnumbered]{algorithm2e} 
\usepackage{todonotes} 
\usepackage{float}
\usepackage{cite}

\newcommand{\R}{\mathbb{R}} 
\newcommand{\rc}{R_c} 

\newcommand{\N}{N} 

\newcommand{\veh}{Q} 
\newcommand{\intr}{I} 
\newcommand{\state}{x} 
\newcommand{\ctrl}{u} 
\newcommand{\dstb}{d} 
\newcommand{\pos}{p} 
\newcommand{\npos}{h} 

\newcommand{\fdyn}{f} 
\newcommand{\cset}{\mathcal{U}} 
\newcommand{\cfset}{\mathbb{U}} 
\newcommand{\dset}{\mathcal{D}} 
\newcommand{\dfset}{\mathbb{D}} 
\newcommand{\obsset}{\mathcal{G}} 
\newcommand{\dz}{\mathcal{Z}} 

\newcommand{\valfunc}{V} 
\newcommand{\valfuncfwd}{W} 
\newcommand{\brs}{\mathcal{V}} 
\newcommand{\frs}{\mathcal{W}} 
\newcommand{\pfrs}{\mathcal{P}} 
\newcommand{\targetset}{\mathcal{L}} 
\newcommand{\ham}{H} 
\newcommand{\fc}{l} 
\newcommand{\ic}{l} 
\newcommand{\obsfunc}{g} 
\newcommand{\costate}{\lambda}

\newcommand{\disckernel}{\Omega} 

\newcommand{\ldt}{t^\text{LDT}} 
\newcommand{\sta}{t^\text{STA}} 
\newcommand{\ioset}{\mathcal{O}} 
\newcommand{\boset}{\mathcal{B}} 
\newcommand{\soset}{\ioset^\text{static}} 
\newcommand{\iat}{t^\text{IAT}} 

\newcommand{\tsa}{\underline{t}} 
\newcommand{\tea}{\bar{t}} 

\newcommand{\errorbound}{\mathcal{E}} 
\newcommand{\tracklaw}{\kappa} 

\newtheorem{assumption}{Assumption}
\newtheorem{alg}{Algorithm}
\newtheorem{remark}{Remark}

\title{\LARGE \bf Robust Sequential Path Planning Under Disturbances and Adversarial Intruder}

\author{Mo Chen, Somil Bansal, Jaime F. Fisac, Claire J. Tomlin
\thanks{This work has been supported in part by NSF under CPS:ActionWebs (CNS-931843), by ONR under the HUNT (N0014-08-0696) and SMARTS (N00014-09-1-1051) MURIs and by grant N00014-12-1-0609, by AFOSR under the CHASE MURI (FA9550-10-1-0567). The research of M. Chen and J. F. Fisac have received funding from the ``NSERC'' program and ``la Caixa" Foundation, respectively.}
\thanks{All authors are with the Department of Electrical Engineering and Computer Sciences, University of California, Berkeley. \{mochen72, somil, jfisac, tomlin\}@eecs.berkeley.edu}
}

\begin{document}
\maketitle
\thispagestyle{empty}
\pagestyle{empty}

\begin{abstract}
Provably safe and scalable multi-vehicle path planning is an important and urgent problem due to the expected increase of automation in civilian airspace in the near future. Although this problem has been studied in the past, there has not been a method that guarantees both goal satisfaction and safety for vehicles with general nonlinear dynamics while taking into account disturbances and potential adversarial agents, to the best of our knowledge. Hamilton-Jacobi (HJ) reachability is the ideal tool for guaranteeing goal satisfaction and safety under such scenarios, and has been successfully applied to many small-scale problems. However, a direct application of HJ reachability in most cases becomes intractable when there are more than two vehicles due to the exponentially scaling computational complexity with respect to system dimension. In this paper, we take advantage of the guarantees HJ reachability provides, and eliminate the computation burden by assigning a strict priority ordering to the vehicles under consideration. Under this sequential path planning (SPP) scheme, vehicles reserve ``space-time'' portions in the airspace, and the space-time portions guarantee dynamic feasibility, collision avoidance, and optimality of the paths given the priority ordering. With a computation complexity that scales quadratically when accounting for both disturbances and an intruder, and \textit{linearly} when accounting for only disturbances, SPP can tractably solve the multi-vehicle path planning problem for vehicles with general nonlinear dynamics in a practical setting. We demonstrate our theory in representative simulations.
\end{abstract}


\section{Introduction}
Recently, there has been an immense surge of interest in the use of unmanned aerial systems (UASs) in urban environments. UASs have great potential in civil applications such as package delivery, aerial surveillance, disaster response, among many others \cite{Tice91, Debusk10, Amazon16, AUVSI16, BBC16}. Unlike previous uses of UASs for military purposes, civil applications will involve unmanned aerial vehicles (UAVs) flying in urban environments, potentially in close proximity of humans, other UAVs, and other important assets. As a result, government agencies such as the Federal Aviation Administration (FAA) and National Aeronautics and Space Administration (NASA) of the United States are urgently trying to develop new scalable ways to organize an airspace in which potentially thousands of UAVs can fly simultaneously in the same region \cite{FAA13, Kopardekar16}.

One essential problem that needs to be addressed is the safe multi-vehicle path planning problem: how a group of vehicles in the same vicinity can reach their destinations while avoiding situations which are considered dangerous, such as collisions. In many previous studies that address this problem, specific control strategies for the vehicles are assumed, and approaches such as those involving induced velocity obstacles \cite{Fiorini98, Chasparis05, Vandenberg08,Wu2012} and involving virtual potential fields to maintain collision \cite{Olfati-Saber2002, Chuang07} have been used. Other analyses of multi-vehicle systems include methods for real-time trajectory generation \cite{Feng-LiLian2002}, for path planning for vehicles with linear dynamics in the presence of obstacles with known motion \cite{Ahmadzadeh2009}, and for cooperative path planning via waypoints which do not account for vehicle dynamics \cite{Bellingham}. Other related work include those which consider only the collision avoidance problem without path planning. These results include those that assume the system has a linear model \cite{Beard2003, Schouwenaars2004, Stipanovic2007}, rely on a linearization of the system model \cite{Massink2001, Althoff2011}, assume a simple positional state space \cite{Lin2015}, and many others \cite{Lalish2008, Hoffmann2008, Chen2016}.

However, the capability to flexibly plan provably safe and dynamically feasible trajectories without making strong assumptions on the vehicles' dynamics and other vehicles' motion is essential for dense groups of UAVs to safely fly in each other's vicinity. In addition, in a practical setting, any path planning scheme that also addresses collision avoidance must guarantee both goal satisfaction and safety of UAVs despite disturbances such as weather effects and communication faults \cite{Kopardekar16}. Furthermore, unexpected scenarios such as UAV malfunctions or even UAVs with malicious intent need to be accounted for.

The problem of trajectory planning and collision avoidance under disturbances in safety-critical systems has been studied using Hamilton-Jacobi (HJ) reachability analysis, which provides guarantees on goal satisfaction and safety of optimal system trajectories \cite{Barron90, Mitchell05, Bokanowski10, Bokanowski11, Margellos11, Fisac15}. Reachability-based methods are particular suitable in the context of UAVs because of the hard guarantees that are provided. In reachability analysis, one computes the reachable set, defined as the set of states from which the system can be driven to a target set. Many numerical tools are available for computing various definitions of reachable sets \cite{Sethian96, Osher02, Mitchell02, Mitchell07b}, and reachability analysis has been successfully used in applications involving systems with no more than two vehicles, such as pairwise collision avoidance \cite{Mitchell05}, automated in-flight refueling \cite{Ding08}, and many others \cite{Huang11, Bayen07}.

One of the main challenges of managing the next generation of airspace is the density of vehicles that needs to be accommodated \cite{Kopardekar16}. Such a large-scale system has a high-dimensional joint state space, making a direct application of dynamic programming-based approaches such as reachability analysis intractable. In particular, reachable set computations involve solving a HJ partial differential equation (PDE) or variational inequality (VI) on a grid representing a discretization of the state space, causing computational complexity to scale \textit{exponentially} with system dimension.

\subsection{Contributions and Outline}

In this paper, we propose the sequential path planning (SPP) method to tackle the multi-vehicle path planning problem. Our approach is similar to the approaches of \cite{Erdmann1987, VandenBerg2005}, but provides hard guarantees on both the goal satisfaction and safety of all vehicles even in the presence of disturbances and a single intruder vehicle that could potentially be adversarial. In addition, our method scales only \textit{linearly} with the number of vehicles when there is no intruder, and \textit{quadratically} with the number of vehicles when there is a single intruder. On a high level, the SPP method assigns a strict priority ordering to the vehicles under consideration. Higher-priority vehicles plan their paths without taking into account the lower-priority vehicles. Lower-priority vehicles treat higher-priority vehicles as moving obstacles. Under this assumption, time-varying formulations of reachability \cite{Bokanowski11, Fisac15} can be used to obtain the optimal and provably safe paths for each vehicle, starting from the highest-priority vehicle. Thus, the curse of dimensionality is overcome for the multi-vehicle path planning problem at the cost of a mild structural assumption. 

In a sense, the SPP method reserves a portion of ``space-time'' in the airspace for each vehicle. The reserved space-time portion is recorded so that lower-priority vehicles can take it into account. Besides planning around the reserved space-time portions of higher-priority vehicles, no other communication between the vehicles is needed at execution time, even when disturbances and an intruder are present.

In the absence of disturbances and intruders, and assuming each vehicle has perfect information about other vehicles' positions, each vehicle may plan and commit to an exact trajectory, with the reserved space-time being the collision set around the trajectory at every point in time. This basic concept of SPP is formally presented in Section \ref{sec:basic}.

When the vehicles are affected by disturbances, exact trajectories cannot be known \textit{a priori}, and thus the basic SPP algorithm cannot be directly applied. Fortunately, reachability analysis allows us to determine, at no additional computation cost, all possible states of each vehicle over time under the worst-case disturbance, given a control strategy. In addition, we can also determine suitable portions of space-time for each vehicle depending on the available information about the control strategies of higher-priority vehicles. SPP under disturbances and three different assumptions on the information available about the control strategy of other vehicles is formally presented in Section \ref{sec:incomp}.

In scenarios where there could potentially be single, possibly adversarial intruder in the airspace, each vehicle needs extra space around other vehicles in order to be able to perform avoidance maneuvers. Assuming the intruder may be present for some maximum duration, we use use reachability analysis to determine precisely the amount of space-time needed for each vehicle to be able to avoid the intruder under the presence of disturbances, making our proposed method sufficiently robust to most practical scenarios. SPP in the presence of a single intruder is formally presented in Section \ref{sec:intruder}.

%
%

\section{Problem Formulation \label{sec:formulation}}
Consider $\N$ vehicles which participate in the SPP process and denote these vehicles as the \textit{SPP vehicles} $\veh_i, i = 1, \ldots, \N$. We assume their dynamics are given by

\begin{equation}
\label{eq:dyn}
\begin{aligned}
\dot\state_i &= \fdyn_i(\state_i, \ctrl_i, \dstb_i), t \le \sta_i \\
\ctrl_i &\in \cset_i, \dstb_i \in \dset_i, i = 1 \ldots, \N
\end{aligned}
\end{equation}

\noindent where $\state_i \in \R^{n_i}$ represents the state of vehicle $\veh_i$, $\ctrl_i \in \cset_i$ the control of $\veh_i$, and $\dstb_i \in \dset_i$ the disturbance experienced by $\veh_i$. For convenience, we partition the state $\state_i$ into the position component $\pos_i \in \R^{n_\pos}$ and the non-position component $\npos_i \in \R^{n_i - n_\pos}$: $\state_i = (\pos_i, \npos_i)$.

We assume that the control functions $\ctrl_i(\cdot), \dstb_i(\cdot)$ are drawn from the set of measurable functions\footnote{A function $f:X\to Y$ between two measurable spaces $(X,\Sigma_X)$ and $(Y,\Sigma_Y)$ is said to be measurable if the preimage of a measurable set in $Y$ is a measurable set in $X$, that is: $\forall V\in\Sigma_Y, f^{-1}(V)\in\Sigma_X$, with $\Sigma_X,\Sigma_Y$ $\sigma$-algebras on $X$,$Y$.}. For convenience, we will use the sets $\cfset_i, \dfset_i$ to respectively denote the set of functions from which the control and disturbance functions $\ctrl_i(\cdot), \dstb_i(\cdot)$ are drawn.

We further assume that the flow field $\fdyn_i: \R^{n_i}\times\cset_i\times\dset_i \rightarrow \R^{n_i}$ is uniformly continuous, bounded, and Lipschitz continuous in $\state_i$ for fixed $\ctrl_i$ and $\dstb_i$. With this assumption, given $\ctrl_i(\cdot) \in \cfset_i, \dstb_i(\cdot) \in \dfset_i$, there exists a unique trajectory solving \eqref{eq:dyn} \cite{EarlA.Coddington1955}. 


In addition, we assume that the disturbances $\dstb_i(\cdot)$ are drawn the set of non-anticipative strategies \cite{Mitchell05} $\Gamma$, defined as follows:
\begin{equation}
\begin{aligned}
& \Gamma := \{\mathcal{N}: \cfset_i \rightarrow \dfset_i:  \ctrl_i(r) = \hat{\ctrl}_i(r) \text{ a. e. } r\in[t,s] \\
& \Rightarrow \mathcal{N}[\ctrl_i](r) = \mathcal{N}[\hat{\ctrl}_i](r) \text{ a. e. } r\in[t,s]\}
\end{aligned}
\end{equation}

Each vehicle $\veh_i$ has initial state $\state^0_i$, and aims to reach its target $\targetset_i$ by some scheduled time of arrival $\sta_i$. The target in general represents some set of desirable states, for example the destination of $\veh_i$. 
In some situations, we may find that it is infeasible for $\veh_i$ to get to $\targetset_i$ at or before $\sta_i$. Whenever unsure, we may first determine the earliest feasible $\sta_i$ as described in Section \ref{sec:intruder}. 

On its way to $\targetset_i$, $\veh_i$ must avoid a set of static obstacles $\soset_i \subset \R^{n_i}$. The interpretation of $\soset_i$ could be a tall building, a region around an airport, or any set of states that are forbidden for each SPP vehicle.

In addition to the static obstacles, each vehicle $\veh_i$ must also avoid the danger zones with respect to every other vehicle $\veh_j, j\neq i$. The danger zones in general can represent any joint configurations between $\veh_i$ and $\veh_j$ that are considered to be unsafe. In this paper, we define the danger zone of $\veh_i$ with respect to $\veh_j$ to be

\begin{equation}
\dz_{ij} = \{(\state_i, \state_j): \|\pos_i - \pos_j\|_2 \le \rc\}
\end{equation}

\noindent whose interpretation is that $\veh_i$ and $\veh_j$ are considered to be in an unsafe configuration when they are within a distance of $\rc$ of each other. For concreteness, we will call $\dz_{ij}$ the collision set, and if $(\state_i, \state_j) \in \dz_{ij}$, then $\veh_i$ and $\veh_j$ are said to have collided.

Given the set of SPP vehicles, their targets $\targetset_i$, the static obstacles $\soset_i$, and the vehicles' danger zones with respect to each other $\dz_{ij}$, we would like, for each vehicle $\veh_i$, to synthesize a controller which guarantees that $\veh_i$ reaches its target $\targetset_i$ at or before the scheduled time of arrival $\sta_i$, while avoiding the static obstacles $\soset_i$ as well as the danger zones with respect to all other vehicles $\dz_{ij}, j\neq i$. In addition, we would like to obtain the latest departure time $\ldt_i$ such that $\veh_i$ can still arrive at $\targetset_i$ on time.

In general, the above optimal path planning problem must be solved in the joint space of all $\N$ SPP vehicles. However, due to the high joint dimensionality, a direct dynamic programming-based solution is intractable. Therefore, we propose to assign a priority to each vehicle, and perform SPP given the assigned priorities. Without loss of generality, let $\veh_j$ have a higher priority than $\veh_i$ if $j<i$. Under the SPP scheme, higher-priority vehicles can ignore the presence of lower-priority vehicles, and perform path planning without taking into account the lower-priority vehicles' danger zones. A lower-priority vehicle $\veh_i$, on the other hand, must ensure that it does not enter the danger zones of the higher-priority vehicles $\veh_j, j<i$; each higher-priority vehicle $\veh_j$ induces a set of time-varying obstacles $\ioset_i^j(t)$, which represents the possible states of $\veh_i$ such that a collision between $\veh_i$ and $\veh_j$ could occur.

It is straight-forward to see that if each vehicle $\veh_i$ is able to plan a trajectory that takes it to $\targetset_i$ while avoiding the static obstacles $\soset_i$ and the danger zones of \textit{higher-priority vehicles} $\veh_j, j<i$, then the set of SPP vehicles $\veh_i, i=1,\ldots,\N$ would all be able to reach their targets safely. With the SPP scheme, the additional structure provided by the vehicle priorities allows us to reduce the complexity of the joint path planning problem. As we will see, under the SPP scheme, path planning can be done sequentially in descending order of vehicle priority in the state space of only a single vehicle. Thus, SPP provides a solution whose complexity scales linearly with the number of vehicles in the presence of disturbances, as opposed to exponentially with a direct application of dynamic programming approaches. In the presence of a single intruder, the computation complexity scaling becomes quadratic.

In the following sections, we will explore SPP under different assumptions. We begin with the basic SPP algorithm in which disturbances are ignored and perfect information of vehicles' positions is assumed. This simplification allows us to clearly establish the basic SPP algorithm. Next, we show how the basic SPP approach can be made robust to disturbances as well as an imperfect knowledge of other vehicles' positions. Finally, we further robustify the SPP approach by considering how the set of SPP vehicles may respond to the presence of an intruder vehicle which may be adversarial. All of our methods use time-varying reachability analysis to provide goal satisfaction and safety guarantees.
\section{Time-Varying Reachability Background \label{sec:HJIVI}}
We will be using reachability analysis to compute either a backward reachable set (BRS) $\brs$, a forward reachable set (FRS) $\frs$, or a sequence of BRSs and FRSs, given some target set $\targetset$, time-varying obstacle $\obsset(t)$, and the Hamiltonian function $\ham$ which captures the system dynamics as well as the roles of the control and disturbance. The BRS $\brs$ in a time interval $[t, t_f]$ or FRS $\frs$ in a time interval $[t_0, t]$ will be denoted by

\begin{equation}
\begin{aligned}
\brs(t, t_f) &\quad \text{ (backward reachable set)}\\
\frs(t_0, t) &\quad \text{ (forward reachable set)}
\end{aligned}
\end{equation}

In the SPP scheme, a lower-priority vehicle must avoid a set of moving obstacles on its way to the target. Several formulations of reachability are able to perform optimal path planning with hard guarantees on safety and performance under disturbances in such a scenario \cite{Bokanowski11, Fisac15}. For our application in SPP, we utilize the time-varying formulation in \cite{Fisac15}, which accounts for the time-varying nature of systems without requiring augmentation of the state space with the time variable. In the formulation in \cite{Fisac15}, a BRS is computed by solving the following \textit{final value} double-obstacle HJ VI:

\begin{equation}
\label{eq:HJIVI_BRS}
\begin{aligned}
\max \Big\{ \min \{&D_t \valfunc(t, \state) + \ham(t, \state, \nabla \valfunc(t, \state)), \fc(\state) - \valfunc(t, \state) \}, \\
&-\obsfunc(t, \state) - \valfunc(t, \state) \Big\} = 0, \quad t \le t_f \\
&\valfunc(t_f, \state) = \max\{\fc(\state), -\obsfunc(t_f, \state)\}
\end{aligned}
\end{equation}

In a similar fashion, the FRS is computed by solving the following \textit{initial value} HJ PDE:

\begin{equation}
\label{eq:HJIVI_FRS}
\begin{aligned}
D_t \valfuncfwd(t, \state) + &\ham(t, \state, \nabla \valfuncfwd(t, \state)) = 0 , \quad t \ge t_0  \\
&\valfuncfwd(t_0, \state) = \max\{\fc(\state), -\obsfunc(t_0, \state)\}
\end{aligned}
\end{equation}

In both \eqref{eq:HJIVI_BRS} and \eqref{eq:HJIVI_FRS}, the function $\ic(\state)$ is the implicit surface function representing the target set $\targetset = \{\state: \ic(\state) \le 0\}$. Similarly, the function $\obsfunc(t, \state)$ is the implicit surface function representing the time-varying obstacles $\obsset(t) = \{\state: \obsfunc(t,\state)\le 0\}$. The BRS $\brs(t, t_f)$ and FRS $\frs(t_0, t)$ are given by

\begin{equation}
\label{eq:implicitValfuncs}
\begin{aligned}
\brs(t, t_f) &= \{\state: \valfunc(t, \state) \le 0\} \\
\frs(t_0, t) &= \{\state: \valfuncfwd(t, \state) \le 0 \}
\end{aligned}
\end{equation}

Some of the reachability computations will not involve an obstacle set $\obsset(t)$, in which case we can simply set $\obsfunc(t, \state) \equiv \infty$ which effectively means that the outside maximum is ignored in \eqref{eq:HJIVI_BRS}. Also, note that unlike in \eqref{eq:HJIVI_BRS}, there is no inner minimization in \eqref{eq:HJIVI_FRS}. As we will see later, we will be using the BRS to determine all states that can reach some target set \textit{within the time horizon} $[t,t_f]$, whereas we will be using the FRS to determine where a vehicle could be \textit{at some particular time} $t$. In addition, \eqref{eq:HJIVI_FRS} has no outer maximum, since the FRSs that we will compute will not involve any obstacles.

%

The Hamiltonian, $\ham(t, \state, \nabla \valfunc(t,\state))$, depends on the system dynamics, and the role of control and disturbance. Whenever $\ham$ does not depend explicit on $t$, we will drop the argument. In addition, the Hamiltonian is an optimization that produces the optimal control $\ctrl^*(t, \state)$ and optimal disturbance $\dstb^*(t, \state)$, once $\valfunc$ is determined. For BRSs, whenever the existence of a control (``$\exists \ctrl$'') or disturbance is sought, the optimization is a minimum over the set of controls or disturbance. Whenever a BRS characterizes the behavior of the system for all controls (``$\forall \ctrl$'') or disturbances, the optimization is a maximum. We will introduce precise definitions of reachable sets, expressions for the Hamiltonian, expressions for the optimal controls as needed for the many different reachability calculations we use. 
%
%
%
%
%
%

\section{SPP Without Disturbances and With Perfect Information\label{sec:basic}}
In this section, we introduce the basic SPP algorithm assuming that there is no disturbance affecting the vehicles, and that each vehicle knows the exact position of higher-priority vehicles. Although in practice, such assumptions do not hold, the basic SPP algorithm can still serve as a useful approximation in certain situations. In addition, the description of the basic SPP algorithm will introduce the notation needed for describing the subsequent, more realistic versions of SPP. We also show simulation results for the basic SPP algorithm. The majority of the content in this section is taken from \cite{Chen15c}.

\subsection{Theory}
Recall that the SPP vehicles $\veh_i, i=1,\ldots,N$, are each assigned a strict priority, with $\veh_j$ having a higher priority than $\veh_i$ if $j<i$. In the absence of disturbances, we can write the dynamics of the SPP vehicles as

\begin{equation}
\label{eq:dyn_no_dstb}
\begin{aligned}
\dot\state_i &= \fdyn_i(\state_i, \ctrl_i), t \le \sta_i \\
\ctrl_i &\in \cset_i, \qquad i = 1 \ldots, \N
\end{aligned}
\end{equation}


In SPP, each vehicle $\veh_i$ plans the path to its target set $\targetset_i$ while avoiding static obstacles $\soset_i$ and the obstacles $\ioset_i^j(t)$ induced by higher-priority vehicles $\veh_j, j<i$. Path planning is done sequentially starting from the first vehicle and proceeding in descending priority, $\veh_1, \veh_2, \ldots, \veh_{\N}$ so that each of the path planning problems can be done in the state space of only one vehicle. During its path planning process, $\veh_i$ ignores the presence of lower-priority vehicles $\veh_k, k>i$, and induces the obstacles $\ioset_k^i(t)$ for $\veh_k, k>i$.

From the perspective of $\veh_i$, each of the higher-priority vehicles $\veh_j, j<i$ induces a time-varying obstacle denoted $\ioset_i^j(t)$ that $\veh_i$ needs to avoid\footnote{Note that the index $k$ in $\ioset_k^i$ denotes vehicles with lower priority than $\veh_i$, and the index $j$ in $\ioset_i^j(t)$ denotes vehicles with higher priority than $\veh_i$.}. Therefore, each vehicle $\veh_i$ must plan its path to $\targetset_i$ while avoiding the union of all the induced obstacles as well as the static obstacles. Let $\obsset_i(t)$ be the union of all the obstacles that $\veh_i$ must avoid on its way to $\targetset_i$:

\begin{equation}
\label{eq:obsseti}
\obsset_i(t)  = \soset_i \cup \bigcup_{j=1}^{i-1} \ioset_i^j(t)
\end{equation}

With full position information of higher priority vehicles, the obstacle induced for $\veh_i$ by $\veh_j$ is simply

\begin{equation}
\label{eq:ioset}
\ioset_i^j(t) = \{\state_i: \|\pos_i - \pos_j(t)\|_2 \le \rc \}
\end{equation}

Each higher priority vehicle $\veh_j$ plans its path while ignoring $\veh_i$. Since path planning is done sequentially in descending order or priority, the vehicles $\veh_j, j<i$ would have planned their paths before $\veh_i$ does. Thus, in the absence of disturbances, $\pos_j(t)$ is \textit{a priori} known, and therefore $\ioset_i^j(t), j<i$ are known, deterministic moving obstacles, which means that $\obsset_i(t)$ is also known and deterministic. Therefore, the path planning problem for $\veh_i$ can be solved by first computing the BRS $\brs_i^\text{basic}(t, \sta_i)$, defined as follows:

\begin{equation}
\label{eq:BRS_basic}
\begin{aligned}
\brs_i^\text{basic}(t, \sta_i) = & \{y: \exists \ctrl_i(\cdot) \in \cfset_i, \state_i(\cdot) \text{ satisfies \eqref{eq:dyn_no_dstb}}, \\
& \forall s \in [t, \sta_i],\state_i(s) \notin \obsset_i(s), \\
& \exists s \in [t, \sta_i], \state_i(s) \in \targetset_i, \state_i(t) = y\}
\end{aligned}
\end{equation}

The BRS $\brs(t, \sta_i)$ can be obtained by solving \eqref{eq:HJIVI_BRS} with $\targetset = \targetset_i$, $\obsset(t) = \obsset_i(t)$, and the Hamiltonian 

\begin{equation}
\label{eq:basicham}
\ham_i^\text{basic}(\state_i, \costate) = \min_{\ctrl_i\in\cset_i} \costate \cdot \fdyn_i(\state_i, \ctrl_i)
\end{equation}

The optimal control for reaching $\targetset_i$ while avoiding $\obsset_i(t)$ is then given by

\begin{equation}
\label{eq:basicOptCtrl}
\ctrl_i^\text{basic}(t, \state_i) = \arg \min_{\ctrl_i\in\cset_i} \costate \cdot \fdyn_i(\state_i, \ctrl_i)
\end{equation}

\noindent from which the trajectory $\state_i(\cdot)$ can be computed by integrating the system dynamics, which in this case are given by \eqref{eq:dyn_no_dstb}. In addition, the latest departure time $\ldt_i$ can be obtained from the BRS $\brs(t, \sta_i)$ as $\ldt_i = \arg \sup_t \{\state_i^0 \in \brs(t, \sta_i)\}$. In summary, the basic SPP algorithm is given as follows:

\begin{alg}
\label{alg:basic}
\textbf{Basic SPP algorithm}: Suppose we are given initial conditions $\state_i^0$, vehicle dynamics \eqref{eq:dyn_no_dstb}, target sets $\targetset_i$, and static obstacles $\soset_i, i = 1\ldots, \N$. For each $i$ in ascending order starting from $i=1$ (which corresponds to descending order of priority),
\begin{enumerate}
\item determine the total obstacle set $\obsset_i(t)$, given in \eqref{eq:obsseti}. In the case $i=1$, $\obsset_i(t) = \soset_i ~ \forall t$;
\item compute the BRS $\brs_i^\text{basic}(t, \sta_i)$ defined in \eqref{eq:BRS_basic}. The latest departure time $\ldt_i$ is then given by $\arg \sup_t \{\state^0_i \in \brs_i^\text{basic}(t, \sta_i)\}$;
\item determine the trajectory $\state_i(\cdot)$ using vehicle dynamics \eqref{eq:dyn_no_dstb}, with the optimal control  $\ctrl_i^\text{basic}(\cdot)$ given by \eqref{eq:basicOptCtrl};
\item given $\state_i(\cdot)$, compute the induced obstacles $\ioset_k^i(t)$ for each $k>i$. In the absence of disturbances, $\ioset_k^i(t)$ is given by \eqref{eq:ioset}.
\end{enumerate}
\end{alg}
\subsection{Numerical Simulations \label{sec:basic_results}}
We now illustrate the basic SPP algorithm using a four-vehicle example. In this example, we will use the following dynamics for each vehicle:

\begin{equation} \label{eqn:NumSimpleDyn}
\begin{aligned}
\dot{\pos}_{x,i} &= v_i \cos\theta_i \\
\dot{\pos}_{y,i} &= v_i \sin\theta_i \\
\dot \theta_i &= \omega_i \\
|\omega_i| &\le \bar\omega
\end{aligned}
\end{equation}

\noindent where $\state_i = (\pos_{x,i}, \pos_{y,i}, \theta_i)$ is the state of vehicle $\veh_i$, $\pos_i = (\pos_{x,i}, \pos_{y,i})$ is the position, $\theta_i$ is the heading, $v_i$ is the speed, and $\omega_i$ is the turn rate. In this example, we assume that the vehicles have constant speed $v_i = 1 ~ \forall i$, and that the control of each vehicle $\veh_i$ is given by $\ctrl_i = \omega_i$ with $|\omega_i| \le \bar\omega = 1 ~ \forall i$. We have chosen these dynamics for clarity of illustration; the SPP algorithm can handle more general systems of the form in which the vehicles have different control bounds and dynamics. 

For this example, the target sets $\targetset_i$ of the vehicles are circles of radius $r$ in the position space; each vehicle is trying to reach some desired set of positions. In terms of the state space $\state_i$, the target sets are defined as

\begin{equation}
\label{eq:target_sim}
\targetset_i = \{\state_i: \|\pos_i - c_i\|_2 \le r\}
\end{equation}

\noindent where $c_i$ are centers of the target circles. For the simulation of the basic SPP algorithm, we used $r = 0.1$. The vehicles have target centers $c_i$, initial conditions $\state_i^0$, and scheduled times of arrivals $\sta_i$ as follows:

\begin{equation} \label{eqn:NumIC}
\begin{aligned}
c_1 = (0.7, 0.2), \quad& \state_1^0 = (-0.5, 0, 0), \quad & \sta_1 = 0 \\
c_2 = (-0.7, 0.2), \quad& \state_2^0 = (0.5, 0, \pi), \quad & \sta_2 = 0.2 \\
c_3 = (0.7, -0.7), \quad& \state_3^0 = \left(-0.6, 0.6, 7\pi/4\right), \quad & \sta_3 = 0.4\\
c_4 = (-0.7, -0.7), \quad & \state_4^0 = \left(0.6, 0.6, 5\pi/4\right), \quad & \sta_4 = 0.6
\end{aligned}
\end{equation}

\begin{figure}
	\centering
	\includegraphics[width=\columnwidth]{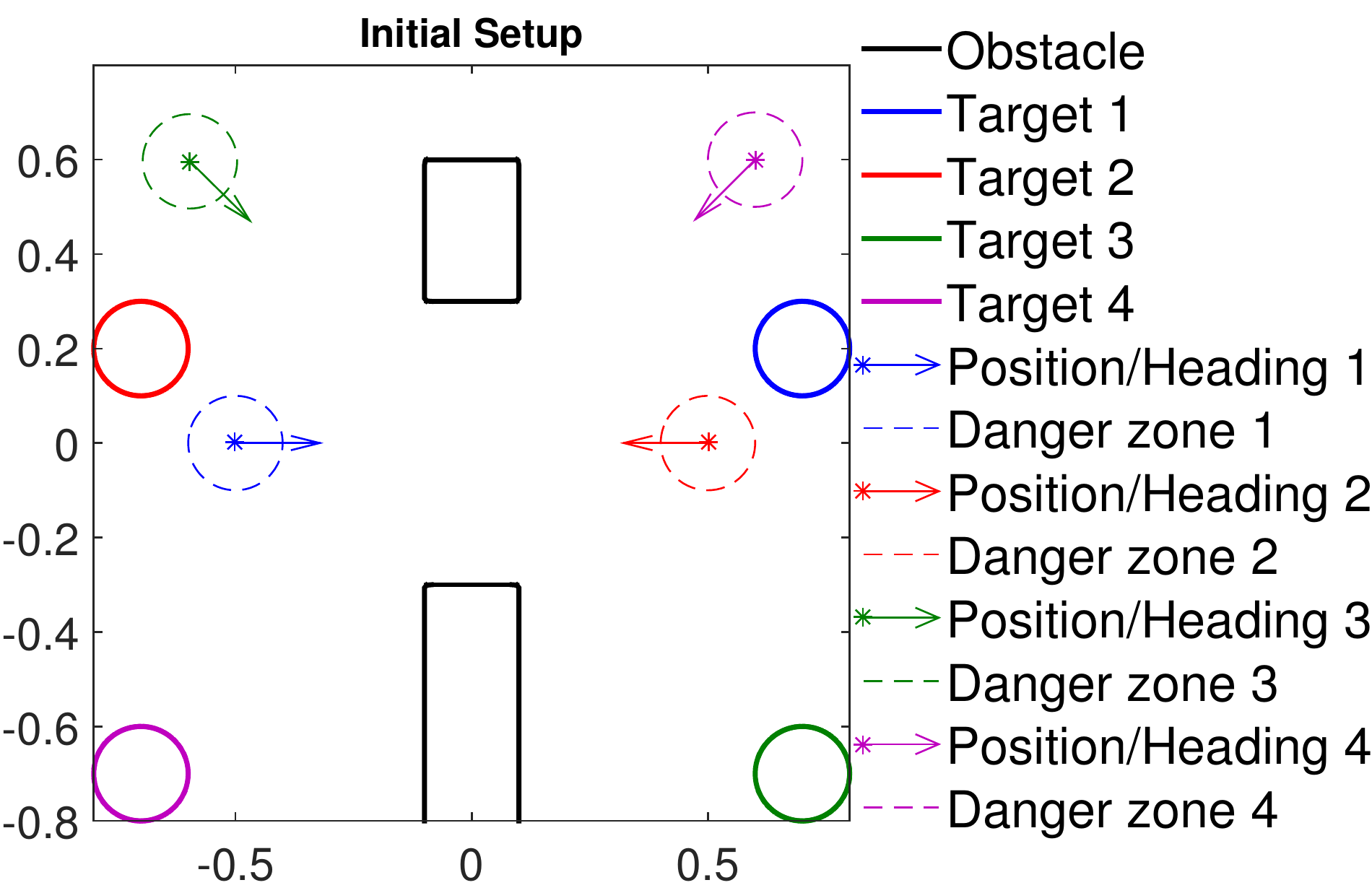}
	\caption{Initial configuration of the four-vehicle example.}
	\label{fig:dubins_ic}
\end{figure}

The setup for this example is shown in Fig. \ref{fig:dubins_ic}, which also shows the static obstacles as the black rectangles around the center of the domain.

The joint state space of this four-vehicle system is twelve-dimensional (12D), making the joint path planning and collision avoidance problem intractable for direct analysis using HJ reachability. Therefore, we apply the SPP algorithm described in Algorithm \ref{alg:basic} and repeatedly solve the double-obstacle HJ VI in \eqref{eq:HJIVI_BRS} to obtain the optimal control for each vehicle to reach its target while avoiding higher-priority vehicles. In addition, due to the flexibility of the HJ VI with respect to time-varying systems, the different scheduled times of arrival $\sta_i$ can be trivially incorporated. 

\begin{figure}
	\centering
	\includegraphics[width=\columnwidth]{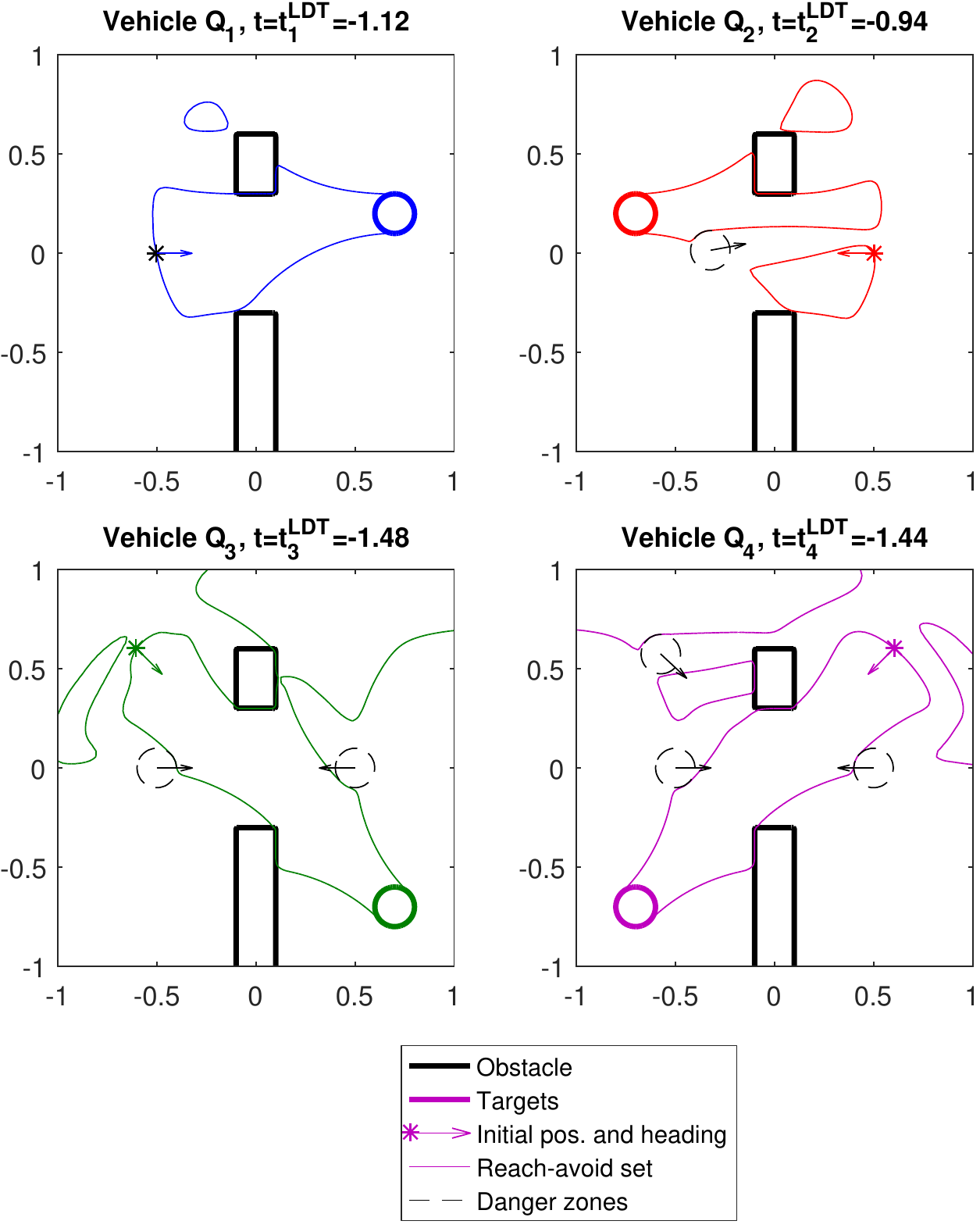}
	\caption{BRSs at $t=\ldt_i$ for vehicles $1,2,3,4$, sliced at initial headings $\theta_i^0$. Black arrows indicate direction of obstacle motion. Due to the turn rate constraint, the presence of static obstacles $\soset_i$ and time-varying obstacles induced by higher-priority vehicles $\ioset_i^j(t)$ carve ``channels" in the BRS, dividing it up into multiple ``islands".}
	\label{fig:dubins_reach_all}
\end{figure}

\begin{figure}
	\centering
	\includegraphics[width=\columnwidth]{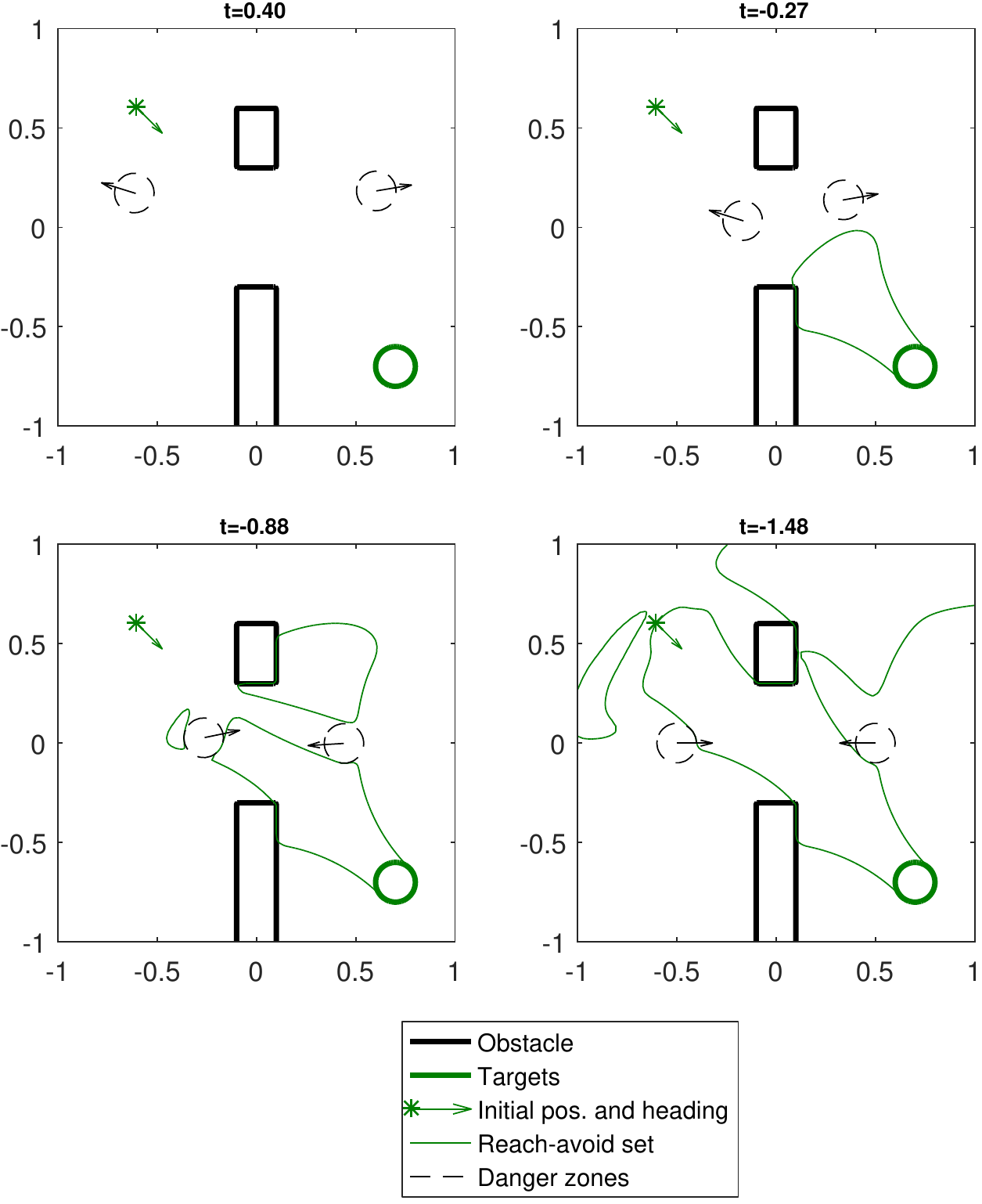}
	\caption{Time evolution of the BRS for vehicle $\veh_3$, sliced at its initial heading $\theta_3^0=\frac{7\pi}{4}$. Black arrows indicate direction of obstacle motion. Top row: the BRS grows unobstructed by obstacles. Bottom row: the static obstacles $\soset_i$ and the induced obstacles $\ioset_3^1,\ioset_3^2$, carve out ``channels" in the BRS.}
	\label{fig:dubins_reach_3}
\end{figure}

\begin{figure}
	\centering
	\includegraphics[width=\columnwidth]{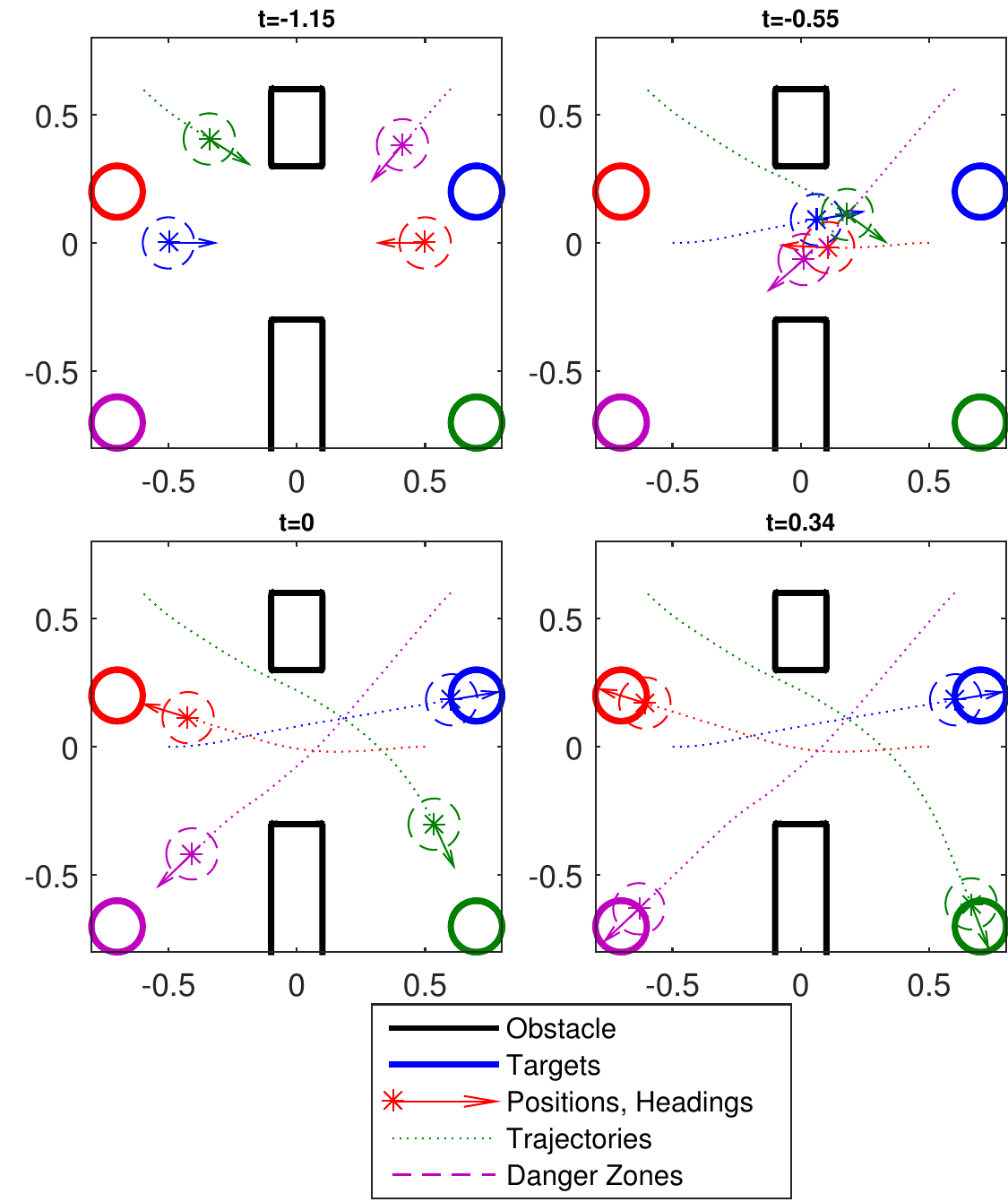}
	\caption{The planned trajectories of the four vehicles. Top left: only vehicles $\veh_3$ (green) and $\veh_4$ (purple) have started moving, showing $\ldt_i$ is not common across the vehicles. Top right: all vehicles have come within very close proximity, but none is in the danger zone of another. Bottom left: vehicle $\veh_1$ (blue) arrives at $\targetset_1$ at $t=0$. Bottom right: all vehicles have reached their destination, some ahead of $\sta_i$.}
	\label{fig:dubins_result}
\end{figure}

Fig. \ref{fig:dubins_reach_all}, \ref{fig:dubins_reach_3}, and \ref{fig:dubins_result} show the simulation results. Since the state space of each vehicle is 3D, each of the BRSs $\brs_i^\text{basic}(t, \sta_i)$ is also 3D. To visualize the results, we slice the BRSs at the initial heading angles $\theta_i^0$. Fig. \ref{fig:dubins_reach_all} shows the 2D BRS slices for each vehicle at its latest departure times $\ldt_1=-1.12, \ldt_2=-0.94,\ldt_3=-1.48,\ldt_4=-1.44$ determined from our method. The obstacles in the domain $\soset_i$ and the obstacles induced by higher-priority vehicles $\ioset_i^j(t)$ inhibit the evolution of the BRSs, carving out thin ``channels" that separate the BRSs into different ``islands". One can see how these ``channels'' and ``islands'' form by examining the time evolution of the BRS, shown in Fig. \ref{fig:dubins_reach_3} for vehicle $\veh_3$. 

Finally, Fig. \ref{fig:dubins_result} shows the resulting trajectories of the four vehicles. Most interestingly, the subplot labeled $t=-0.55$ shows all four vehicles in close proximity without collision: each vehicle is outside of the danger zone of all other vehicles (although the danger zones may overlap). This close proximity is an indication of the optimality of the basic SPP algorithm given the assigned priority ordering. Since no disturbances are present, getting as close to other vehicles' danger zones as possible without entering the danger zones intuitively results in short transit times.

The actual arrival times of vehicles $\veh_i,i=1,2,3,4$ are $0, 0.19, 0.34, 0.31$, respectively. It is interesting to note that for some vehicles, the actual arrival times are earlier than the scheduled times of arrivals $\sta_i$. This is because in order to arrive at the target by $\sta_i$, these vehicles must depart early enough to avoid major delays resulting from the induced obstacles of other vehicles; these delays would have led to a late arrival if vehicle $\veh_i$ departed after $\ldt_i$.

\section{SPP With Disturbances and Incomplete Information \label{sec:incomp}}
Disturbances and incomplete information significantly complicate the SPP scheme. The main difference is that the vehicle dynamics satisfy \eqref{eq:dyn} as opposed to \eqref{eq:dyn_no_dstb}. Committing to exact trajectories is therefore no longer possible, since the disturbance $\dstb_i(\cdot)$ is \textit{a priori} unknown. Thus, the induced obstacles $\ioset_i^j(t)$ are no longer just the danger zones centered around positions. 

\subsection{Theory}
We present three methods to address the above issues. The methods differ in terms of control policy information that is known to a lower-priority vehicle, and have their relative advantages and disadvantages depending on the situation. The three methods are as follows:
\begin{itemize}
\item \textbf{Centralized control}: A specific control strategy is enforced upon a vehicle; this can be achieved, for example, by some central agent such as an air traffic controller.
\item \textbf{Least restrictive control}: A vehicle is required to arrive at its target on time, but has no other restrictions on its control policy. When the control policy of a vehicle is unknown, but its timely arrive at its target can be assumed, the least restrictive control can be safely assumed by lower-priority vehicles.
\item \textbf{Robust trajectory tracking}: A vehicle declares a nominal trajectory which can be robustly tracked under disturbances.
\end{itemize}

In general, the above methods can be used in combination in a single path planning problem, with each vehicle independently having different control policies. Lower-priority vehicles would then plan their paths while taking into account the control policy information known for each higher-priority vehicle. For clarity, we will present each method as if all vehicles are using the same method of path planning.

In addition, for simplicity of explanation, we will assume that no static obstacles exist. In the situations where static obstacles do exist, the time-varying obstacles $\obsset_i(t)$ simply become the union of the induced obstacles $\ioset_i^j(t)$ in \eqref{eq:ioset} and the static obstacles. The material in this section is taken partially from \cite{Bansal2017}.

\subsubsection{Centralized Control\label{sec:cc}}
The highest-priority vehicle $\veh_1$ first plans its path by computing the BRS (with $i=1$)
\begin{equation}
\label{eq:BRS}
\begin{aligned}
\brs_i^\text{dstb}(t, \sta_i) = & \{y: \exists \ctrl_i(\cdot) \in \cfset_i, \forall \dstb_i(\cdot) \in \dfset_i, \state_i(\cdot) \text{ satisfies \eqref{eq:dyn}},\\
& \forall s \in [t, \sta_i], \state_i(s) \notin \obsset_i(s), \state_i(t) = y\\
& \exists s \in [t, \sta_i], \state_i(s) \in \targetset_i\}
\end{aligned}
\end{equation}

Since we have assumed no static obstacles exist, we have that for $\veh_1, \obsset_1(s)=\emptyset ~ \forall s \le \sta_i$, and thus the above BRS is well-defined. This BRS can be computed by solving the HJ VI \eqref{eq:HJIVI_BRS} with the following Hamiltonian:

\begin{equation}
\ham_i^\text{dstb}\left(\state_i, \costate\right) = \min_{\ctrl_i \in \cset_i} \max_{\dstb_i \in \dset_i} \costate \cdot \fdyn_i(\state_i, \ctrl_i, \dstb_i)
\end{equation}

From the BRS, we can obtain the optimal control

\begin{equation}
\label{eq:opt_ctrl_i}
\ctrl_i^\text{dstb}(t,\state_i) =  \arg \min_{\ctrl_i \in \cset_i} \max_{\dstb_i \in \dset_i} \costate \cdot \fdyn_i(\state_i, \ctrl_i, \dstb_i)
\end{equation}

Here, as well as in the other two methods, the latest departure time $\ldt_i$ is then given by $\arg \sup_t \state_{i}^0 \in \brs_i^\text{dstb}(t, \sta_i)$.

If there is a central agent directly controlling each of the $N$ vehicles, then the control law of each vehicle can be enforced. In this case, lower-priority vehicles can safely assume that higher-priority vehicles are applying the enforced control law. In particular, the optimal controller for getting to the target, $\ctrl^\text{dstb}_i(t, \state_i)$, can be enforced. In this case, the dynamics of each vehicle becomes 

\begin{equation}
\label{eq:dyn_cc}
\begin{aligned}
\dot \state_i &= \fdyn^\text{cc}_i (t, \state_i, \dstb_i) = \fdyn_i(\state_i, \ctrl^\text{dstb}_i(t,\state_i), \dstb_i) \\
\dstb_i &\in \dset_i, \quad i = 1,\ldots, N, \quad t \in [\ldt_i, \sta_i]
\end{aligned}
\end{equation}

\noindent where $\ctrl_i$ no longer appears explicitly in the dynamics.

From the perspective of a lower-priority vehicle $\veh_i$, a higher-priority vehicle $\veh_j, j < i$ induces a time-varying obstacle that represents the positions that could possibly be within the collision radius $\rc$ of $\veh_j$ under the dynamics $\fdyn^\text{cc}_j(t, \state_j, \dstb_j)$. Determining this obstacle involves computing an FRS of $\veh_j$ starting from\footnote{In practice, we define the target set to be a small region around the vehicle's initial state for computational reasons.} $\state_j(\ldt_j) = \state_{j}^0$. The FRS $\frs_j^\text{cc}(\ldt_j, t)$ is defined as follows:

\begin{equation}
\label{eq:FRS1}
\begin{aligned}
\frs_j^\text{cc}(\ldt_j, t) = & \{y: \exists \dstb_j(\cdot) \in \dfset_j, \state_j(\cdot) \text{ satisfies \eqref{eq:dyn_cc}},\\
& \state_j(\ldt_j) = \state_{j}^0, \state_j(t) = y\}.
\end{aligned}
\end{equation}

This FRS can be computed using \eqref{eq:HJIVI_FRS} with the Hamiltonian

\begin{equation}
\ham_j^\text{cc}\left(t, \state_j, \costate\right) = \max_{\dstb_j \in \dset_j} \costate \cdot f^\text{cc}_j(t, \state_j, \dstb_j)
\end{equation}

The FRS $\frs_j^\text{cc}(\ldt_j, t)$ represents the set of possible states at time $t$ of a higher-priority vehicle $\veh_j$ given all possible disturbances $\dstb_j(\cdot)$ and given that $\veh_j$ uses the feedback controller $\ctrl_j^\text{dstb}(t, \state_j)$. In order for a lower-priority vehicle $\veh_i$ to guarantee that it does not go within a distance of $\rc$ to $\veh_j$, $\veh_i$ must stay a distance of at least $\rc$ away from the FRS $\frs_j^\text{cc}(\ldt_j, t)$ for all possible values of the non-position states $\npos_j$. This gives the obstacle induced by a higher-priority vehicle $\veh_j$ for a lower-priority vehicle $\veh_i$ as follows:

\begin{equation} \label{eqn:ccObs}
\ioset_i^j(t) = \{\state_i: \exists y \in \pfrs_j(t), \|\pos_i - y\|_2 \le \rc \}
\end{equation}

\noindent where the set $\pfrs_j(t)$ is the set of states in the FRS $\frs_j^\text{cc}(\ldt_j, t)$ projected onto the states representing position $\pos_j$, and disregarding the non-position dimensions $\npos_j$:

\begin{align} 
\pfrs_j(t) & = \{p_j: \exists \npos_j, (p_j, \npos_j) \in \boset_j(t) \}, \label{eqn:ccObs_help1}\\
\boset_j(t) & = \frs_j^\text{cc}(\ldt_j, t). \label{eqn:ccObs_help2}
\end{align}

Finally, taking the union of the induced obstacles $\ioset_i^j(t)$ as in \eqref{eq:obsseti} gives us the time-varying obstacles $\obsset_i(t)$ needed to define and determine the BRS $\brs_i^\text{dstb}(t, \sta_i)$ in \eqref{eq:BRS}. Repeating this process, all vehicles will be able to plan paths that guarantee the vehicles' timely and safe arrival. The centralized control algorithm can be summarized as follows:
\begin{alg}
\label{alg:cc}
\textbf{Centralized control algorithm}: Given initial conditions $\state_i^0$, vehicle dynamics \eqref{eq:dyn}, target set $\targetset_i$, and static obstacles $\soset_i, i = 1\ldots, \N$, for each $i$,
\begin{enumerate}
\item determine the total obstacle set $\obsset_i(t)$, given in \eqref{eq:obsseti}. In the case $i=1$, $\obsset_i(t) = \soset_i ~ \forall t$;
\item compute the BRS $\brs_i^\text{dstb}(t, \sta_i)$ defined in \eqref{eq:BRS}. The latest departure time $\ldt_i$ is then given by $\arg \sup_t \state^0_i \in \brs_i^\text{dstb}(t, \sta_i)$;
\item compute the optimal control $\ctrl_i^\text{dstb}(t,\state_i)$ corresponding to $\brs_i^\text{dstb}(t, \sta_i)$ given by \eqref{eq:opt_ctrl_i}. Given $\ctrl_i^\text{dstb}(t,\state_i)$, compute the FRS $\frs_i^\text{cc}(\ldt_i, t)$ in \eqref{eq:FRS1};
\item finally, compute the induced obstacles $\ioset_k^i(t)$ for each $k>i$. In the centralized control method, $\ioset_k^i(t)$ is computed using \eqref{eqn:ccObs} where $\pfrs_i(t)$ is given by \eqref{eqn:ccObs_help1}.
\end{enumerate}
\end{alg}
\subsubsection{Least Restrictive Control \label{sec:lrc}}
Here, we again begin with the highest-priority vehicle $\veh_1$ planning its path by computing the BRS $\brs_i^\text{dstb}(t, \sta_i)$ in \eqref{eq:BRS}. However, if there is no centralized controller to enforce the control policy for higher-priority vehicles, weaker assumptions must be made by the lower-priority vehicles to ensure collision avoidance. One reasonable assumption is that all higher-priority vehicles follow the least restrictive control that would take them to their targets. This control would be given by 

\begin{equation}
\label{eq:lrctrl} 
\ctrl_j^\text{lrc}(t, \state_j)\in \begin{cases} \{\ctrl_j^\text{dstb}(t, \state_j) \text{ in } \eqref{eq:opt_ctrl_i}\} \text{ if } \state_j(t)\in \partial \brs_j^\text{dstb}(t, \sta_j), \\
\cset_j  \text{ otherwise}
\end{cases}
\end{equation}

Such a controller allows each higher-priority vehicle to use any controller it desires, except when it is on the boundary of the BRS, $\partial \brs_j^\text{dstb}(t, \sta_j)$, in which case the optimal control $\ctrl_j^\text{dstb}(t, \state_j)$ given by \eqref{eq:opt_ctrl_i} must be used to get to the target safely and on time. This assumption is the weakest assumption that could be made by lower-priority vehicles given that the higher-priority vehicles will get to their targets on time.

Suppose a lower-priority vehicle $\veh_i$ assumes that higher-priority vehicles $\veh_j, j < i$ use the least restrictive control strategy $\ctrl_j^\text{lrc}(t, \state_j)$ in \eqref{eq:lrctrl}. From the perspective of the lower-priority vehicle $\veh_i$, a higher-priority vehicle $\veh_j$ could be in any state that is reachable from $\veh_j$'s initial state $\state_j(\ldt_j) = \state_{j}^0$ and from which the target $\targetset_j$ can be reached. Mathematically, this is defined by the intersection of an FRS $\frs_j^\text{lrc}(\ldt_j, t)$ from the initial state $\state_j(\ldt_j)=\state_{j}^0$ and the BRS $\brs_j^\text{dstb}(t, \sta_j)$ defined in \eqref{eq:BRS} from the target set $\targetset_j$, $\frs_j^\text{lrc}(\ldt_j, t) \cap \brs_j^\text{dstb}(t, \sta_j)$. In this situation, since $\veh_j$ cannot be assumed to be using any particular feedback control, $\frs_j^\text{lrc}(\ldt_j, t)$ is defined as

\begin{equation}
\label{eq:FRS2}
\begin{aligned}
\frs_j^\text{lrc}(\ldt_j, t) = & \{y: \exists \ctrl_j(\cdot)\in\cfset_j, \exists \dstb_j(\cdot) \in \dfset_j, \\
& \state_j(\cdot) \text{ satisfies \eqref{eq:dyn}}, \state_j(\ldt_j) = \state_{j}^0,\\
& \state_j(t) = y\}.
\end{aligned}
\end{equation}

This FRS can be computed by solving \eqref{eq:HJIVI_FRS} with the Hamiltonian

\begin{equation}
\ham_j^\text{lrc}\left(\state_j, \lambda\right) = \max_{\ctrl_j \in \cset_j} \max_{\dstb_j \in \dset_j} \lambda \cdot \fdyn_j(\state_j, \ctrl_j, \dstb_j)
\end{equation}

In turn, the obstacle induced by a higher-priority $\veh_j$ for a lower-priority vehicle $\veh_i$ is as follows:

\begin{equation}
\label{eqn:lrcObs1}
\ioset_i^j(t) = \{\state_i: \exists y \in \pfrs_j(t), \|\pos_i - y\|_2 \le \rc \}, \text{ where}
\end{equation}

\begin{align}
\pfrs_j(t) & = \{p_j: \exists \npos_j, (p_j, \npos_j) \in \boset_j(t) \}, \label{eqn:lrcObs2}\\
\boset_j(t) & = \frs_j^\text{lrc}(\ldt_j, t) \cap \brs_j^\text{dstb}(t, \sta_j). \label{eqn:lrcObs3}
\end{align}

The least restrictive control method can be summarized as follows:
\begin{alg}
\label{alg:lrc}
\textbf{Least restrictive control algorithm}: Given initial conditions $\state_i^0$, vehicle dynamics \eqref{eq:dyn}, target set $\targetset_i$, and static obstacles $\soset_i, i = 1\ldots, \N$, for each $i$,
\begin{enumerate}
\item determine the total obstacle set $\obsset_i(t)$, given in \eqref{eq:obsseti}. In the case $i=1$, $\obsset_i(t) = \soset_i ~ \forall t$;
\item compute the BRS $\brs_i^\text{dstb}(t, \sta_i)$ defined in \eqref{eq:BRS}. The latest departure time $\ldt_i$ is then given by $\arg \sup_t \state_{i}^0 \in \brs_i^\text{dstb}(t, \sta_i)$;
\item compute the FRS $\frs_i^\text{lrc}(t)$ in \eqref{eq:FRS2}. Given $\frs_i^\text{lrc}(\ldt_i, t)$ and $\brs_i^\text{dstb}(t, \sta_i)$, compute the positions that the $\veh_i$ could be in. The set of these positions is given by \eqref{eqn:lrcObs2};
\item compute the induced obstacles $\ioset_k^i(t)$ for each $k>i$ using \eqref{eqn:lrcObs1}.
\end{enumerate}
\end{alg}

\begin{remark}
The centralized control method described in the previous section can be thought of as the ``most restrictive control'' method, in which all vehicles must use the optimal controller at all times, while the least restrictive control method allows vehicles to use any suboptimal controller that allows them to arrive at the target on time. These two methods can be considered two extremes of a spectrum in which varying degrees of optimality is assumed for higher-priority vehicles. Vehicles can also choose a control strategy in the middle of the two extremes, and for example use a control within some range around the optimal control, or use the optimal control unless some condition is met. The induced obstacles and the BRS can then be similarly computed using the corresponding control strategy.
\end{remark}
\subsubsection{Robust Trajectory Tracking\label{sec:rtt}}
Even though it is impossible to commit to and track an exact trajectory in the presence of disturbances, it may still be possible to instead \textit{robustly} track a feasible \textit{nominal} trajectory with a bounded error at all times. If this can be done, then the tracking error bound can be used to determine the induced obstacles. Here, computation is done in two phases: the \textit{planning phase} and the \textit{disturbance rejection phase}. In the planning phase, we compute a nominal trajectory $\state_{r,j}(\cdot)$ that is feasible in the absence of disturbances. In the disturbance rejection phase, we compute a bound on the tracking error.

It is important to note that the planning phase does not make full use of a vehicle's control authority, as some margin is needed to reject unexpected disturbances while tracking the nominal trajectory. Therefore, in this method, planning is done for a reduced control set $\cset^p\subset\cset$. The resulting trajectory reference will not utilize the vehicle's full control capability; additional maneuverability is available at execution time to counteract external disturbances.

In the disturbance rejection phase, we determine the error bound independently of the nominal trajectory. To compute this error bound, we find a robust controlled-invariant set in the joint state space of the vehicle and a tracking reference that may ``maneuver" arbitrarily in the presence of an unknown bounded disturbance. Taking a worst-case approach, the tracking reference can be viewed as a virtual evader vehicle that is optimally avoiding the actual vehicle to enlarge the tracking error. We therefore can model trajectory tracking as a pursuit-evasion game in which the actual vehicle is playing against the coordinated worst-case action of the virtual vehicle and the disturbance. 


Let $\state_j$ and $\state_{r,j}$ denote the states of the actual vehicle $\veh_j$ and the virtual evader, respectively, and define the tracking error $e_j=\state_j-\state_{r,j}$. When the error dynamics are independent of the absolute state as in \eqref{eq:edyn} (and also (7) in \cite{Mitchell05}), we can obtain error dynamics of the form

\begin{equation}
\label{eq:edyn} 
\begin{aligned}
\dot{e_j} &= \fdyn_{e_j}(e_j, \ctrl_j, \ctrl_{r,j},\dstb_j), \\
\ctrl_j &\in \cset_j, \ctrl_{r,j} \in \cset^p_j, \dstb_j \in \dset_j, \quad t \leq 0
\end{aligned}
\end{equation}

To obtain bounds on the tracking error, we first conservatively estimate the error bound around any reference state $\state_{r,j}$, denoted $\errorbound_j$:

\begin{equation} \label{eqn:err}
\errorbound_j = \{e_j: \|\pos_{e_j}\|_2 \le R_{\text{EB}} \}, 
\end{equation}

\noindent where $\pos_{e_j}$ denotes the position coordinates of $e_j$ and $R_{\text{EB}}$ is a design parameter. We next solve a reachability problem with its complement $\errorbound_j^c$, the set of tracking errors violating the error bound, as the target in the space of the error dynamics. From $\errorbound_j^c$, we compute the following BRS:

\begin{equation} \label{eqn:errBound}
\begin{aligned}
\brs^{\text{EB}}_{j}(t, 0) = & \{y: \forall \ctrl_j(\cdot) \in \cfset_j, \exists \ctrl_{r, j}(\cdot) \in \cfset^\pos_j, \exists \dstb_j(\cdot) \in \dfset_i, \\
& e_j(\cdot) \text{ satisfies \eqref{eq:edyn}}, e_j(t) = y, \\
& \exists s \in [t, 0], e_j(s) \in \errorbound_j^c\}, 
\end{aligned}
\end{equation}
where the Hamiltonian to compute the BRS is given by:
\begin{equation}
\begin{aligned}
H^{\text{EB}}_{j}(e_j, \costate) &= \max_{\ctrl_j \in \cset_j} \min_{\ctrl_r \in \cset^\pos_j, \dstb_j \in \dset_j} \costate \cdot \fdyn_{e_j}(e_j, \ctrl_j, \ctrl_{r,j}, \dstb_j).
\end{aligned}
\end{equation}

Letting $t \to -\infty$, we obtain the infinite-horizon control-invariant set $\disckernel_j := \lim_{t \to -\infty} \left( \brs^{\text{EB}}_{j}(t, 0) \right)^c$. If $\disckernel_j$ is nonempty, then the tracking error $e_j$ at flight time is guaranteed to remain within $\disckernel_j \subseteq \errorbound_j$ provided that the vehicle starts inside $\disckernel_j$ and subsequently applies the feedback control law

\begin{equation}
\label{eq:robust_tracking_law}
\tracklaw_j(e_j) = \arg\max_{\ctrl_j \in \cset_j} \min_{\ctrl_r \in\cset^\pos_j, \dstb_j \in \dset_j} \costate \cdot \fdyn_{e_j}(e_j,\ctrl_j,\ctrl_{r, j},\dstb_j).
\end{equation}

The induced obstacles by each higher-priority vehicle $\veh_j$ can thus be obtained by: 
\begin{equation} 
\label{eqn:rttObs}
\begin{aligned}
\ioset_i^j(t) &=  \{\state_i: \exists y \in \pfrs_j(t), \|\pos_i - y\|_2 \le \rc \} \\
\pfrs_j(t) &= \{\pos_j: \exists \npos_j, (\pos_j, \npos_j) \in \boset_j(t)\} \\
\boset_j(t) &= \disckernel_j  + \state_{r,j}(t),
\end{aligned}
\end{equation}

\noindent where the ``$+$'' in \eqref{eqn:rttObs} denotes the Minkowski sum\footnote{The Minkowski sum of sets $A$ and $B$ is the set of all points that are the sum of any point in $A$ and $B$.}. Intuitively, if $\veh_j$ is tracking $\state_{r,j}(t)$, then it will remain within the error bound $\disckernel_j$ around $\state_{r,j}(t) ~\forall t$. This is precisely the set $\pfrs_j(t)$. The induced obstacles can then be obtained by augmenting a danger zone around this set. Finally, we can obtain the total obstacle set $\obsset_i(t)$ using \eqref{eq:obsseti}. 

Since each vehicle $\veh_j$, $j<i$, can only be guaranteed to stay within $\disckernel_j$, we must make sure during the path planning of $\veh_i$ that at any given time, the error bounds of $\veh_i$ and $\veh_j$, $\disckernel_i$ and $\disckernel_j$, do not intersect. This can be done by augmenting the total obstacle set by $\disckernel_i$:

\begin{equation} 
\label{eqn:rttAugObs}
\tilde{\obsset}_i(t) = \obsset_i(t) + \disckernel_i.
\end{equation}

Finally, given $\disckernel_i$, we can guarantee that $\veh_i$ will reach its target $\targetset_i$ if $\disckernel_i \subseteq \targetset_i$; thus, in the path planning phase, we modify $\targetset_i$ to be $\tilde{\targetset}_i := \{\state_i: \disckernel_i + \state_i \subseteq \targetset_i\}$, and compute a BRS, with the control authority $\cset^\pos_i$, that contains the initial state of the vehicle. Mathematically,

\begin{equation}
\label{eq:rttBRS}
\begin{aligned}
\brs_i^\text{rtt}(t, \sta_i) = & \{y: \exists \ctrl_i(\cdot) \in \cfset^p_i, \state_i(\cdot) \text{ satisfies \eqref{eq:dyn_no_dstb}},\\
&\forall s \in [t, \sta_i], \state_i(s) \notin \tilde{\obsset}_i(t), \\
& \exists s \in [t, \sta_i], \state_i(s) \in \tilde{\targetset}_i, \state_i(t) = y\}
\end{aligned}
\end{equation}

The BRS $\brs_i^\text{rtt}(t, \sta_i)$ can be obtained by solving \eqref{eq:HJIVI_BRS} using the Hamiltonian: 
\begin{equation}
\label{eq:RTTham}
\ham_i^\text{rtt}(\state_i, \costate) = \min_{\ctrl_i \in \cset^\pos_i } \costate \cdot \fdyn_i(\state_i, \ctrl_i)
\end{equation}

The corresponding optimal control for reaching $\tilde{\targetset}_i$ is given by:
\begin{equation}
\label{eq:RTTOptCtrl}
\ctrl_i^\text{rtt}(t) = \arg \min_{\ctrl_i \in \cset^\pos_i } \costate \cdot \fdyn_i(\state_i, \ctrl_i).
\end{equation}

The nominal trajectory $\state_{r,i}(\cdot)$ can thus be obtained by using vehicle dynamics \eqref{eq:dyn_no_dstb}, with the optimal control  $\ctrl_i^\text{rtt}(\cdot)$ given by \eqref{eq:RTTOptCtrl}. From the resulting nominal trajectory $\state_{r,i}(\cdot)$, the overall control policy to reach $\targetset_i$ can be obtained via \eqref{eq:robust_tracking_law}. The robust trajectory tracking method can be summarized as follows:

\begin{alg}
\label{alg:rtt}
\textbf{Robust trajectory tracking algorithm}: Given initial conditions $\state_i^0$, vehicle dynamics \eqref{eq:dyn}, target sets $\targetset_i$, and static obstacles $\soset_i, i = 1\ldots, \N$, for each $i$,
\begin{enumerate}
\item determine the total obstacle set $\obsset_i(t)$, given in \eqref{eq:obsseti}. In the case $i=1$, $\obsset_i(t) = \soset_i ~ \forall t$;
\item decide on a reduced control authority $\cset^\pos_i$ for the planning phase, and choose a parameter $R_{\text{EB}}$ to conservatively bound the tracking error;
\item compute the BRS $\brs^{\text{EB}}_{i}(t, 0)$ using \eqref{eqn:errBound} and make sure that $\disckernel_i \neq \emptyset$. Given $R_{\text{EB}}$, the error bound on the tracking error is given by $\disckernel_i$;
\item using $\disckernel_i$, determine the augmented obstacle set $\tilde{\obsset}_i(t)$, given in \eqref{eqn:rttAugObs};
\item compute the BRS $\brs_i^\text{rtt}(t, \sta_i)$ as described in \eqref{eq:rttBRS} using the reduced target set $\tilde{\targetset}_i$, $\tilde{\obsset}_i(t)$ as obstacles, and the control authority $\cset^\pos_i$. The latest departure time $\ldt_i$ is then given by $\arg \sup_t \state^0_i \in \brs_i^\text{rtt}(t, \sta_i)$;
\item compute the nominal trajectory $\state_{r,i}(\cdot)$ for $\veh_i$ in the absence of disturbances, which can be obtained using the vehicle dynamics in \eqref{eq:dyn_no_dstb} and the optimal control given in \eqref{eq:RTTOptCtrl};
\item the induced obstacles $\ioset_k^i(t)$ for each $k>i$ can be computed using $\disckernel_i$ and $\state_{r,i}(\cdot)$ via \eqref{eqn:rttObs}.
\end{enumerate}
\end{alg}

\subsection{Numerical Simulations \label{sec:sim_dstb}}
We demonstrate our proposed methods for accounting for disturbances and incomplete information using a four-vehicle example. Each vehicle has the simple kinematics model in \eqref{eqn:NumSimpleDyn} but with disturbances added to the evolution of each state:
\begin{equation}
\label{eq:dyn_i}
\begin{aligned}
\dot{\pos}_{x,i} &= v_i \cos \theta_i + d_{x,i} \\
\dot{\pos}_{y,i} &= v_i \sin \theta_i + d_{y,i}\\
\dot{\theta}_i &= \omega_i + d_{\theta,i}, \\
\underline{v} & \le v_i \le \bar{v}, |\omega_i| \le \bar{\omega},\\
\|(d_{x,i}, & d_{y,i}) \|_2 \le d_{r}, |d_{\theta,i}| \le \bar{d_{\theta}}
\end{aligned}
\end{equation}

\noindent where $d = (d_{x,i}, d_{y,i}, d_{\theta,i})$ represents $\veh_i$'s disturbances in the three states. The control of $\veh_i$ is $u_i = (v_i, \omega_i)$, where $v_i$ is the speed of $\veh_i$ and $\omega_i$ is the turn rate; both controls have a lower and upper bound. For illustration purposes, we choose $\underline{v} = 0.5, \bar{v} = 1, \bar\omega = 1$; however, our method can easily handle the case in which these inputs differ across vehicles. The disturbance bounds are chosen as $d_r = 0.1, \bar{d_{\theta}} = 0.2$, which correspond to a 10\% uncertainty in the dynamics. 

%

\begin{figure}[H]
  \centering
  \includegraphics[width=\columnwidth]{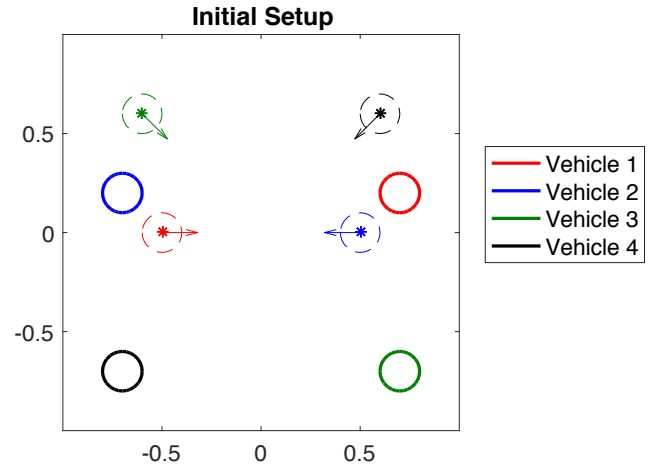}
  \caption{Initial configuration of the four-vehicle example in the presence of disturbances.}
  \label{fig:init_setup_dstb}
\end{figure}

For this example, we have chosen scheduled times of arrival $\sta_i = 0~\forall i$ for simplicity. Each vehicle aims to get to a target set of the form \eqref{eq:target_sim} with target radius $r=0.1$. The vehicles' target centers $c_i$ and initial conditions $\state_i^0$ are given by \eqref{eqn:NumIC}.

These parameters are the same as the example in Section \ref{sec:basic_results}, except that the $\sta_i$ values are the same for all vehicles, and that there are no static obstacles. The problem setup for this example is shown in Fig. \ref{fig:init_setup_dstb}.

With the above parameters, we obtain $\ldt_i, i=1,2,3,4$. Note that even though $\sta_i$ is assumed to be same for all vehicles in this example for simplicity, our method can easily handle the case in which $\sta_i$ is different for each vehicle as we have already shown in Section \ref{sec:basic_results}.

For each proposed method of computing induced obstacles, we show the vehicles' entire trajectories (colored dotted lines), and overlay their positions (colored asterisks) and headings (arrows) at a point in time in which they are in a relatively dense configuration. In all cases, the vehicles are able to avoid each other's danger zones (colored dashed circles) while getting to their target sets in minimum time. In addition, we show the evolution of the BRS over time for $\veh_3$ (green boundaries) as well as the obstacles induced by the higher-priority vehicles (black boundaries).

\begin{figure}[H]
  \centering
  \includegraphics[width=\columnwidth]{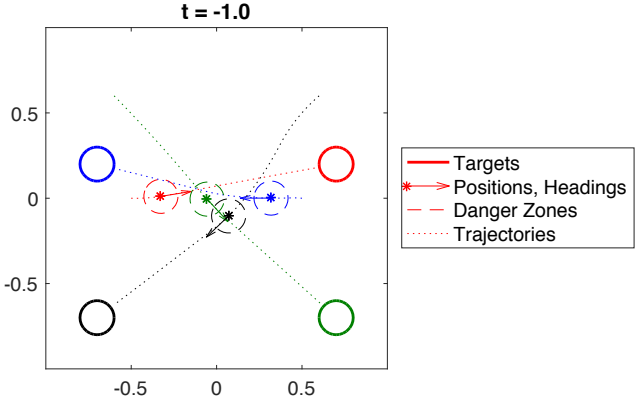}
  \caption{Simulated trajectories in the centralized control method. Since the higher priority vehicles induce relatively small obstacles in this case, vehicles do not deviate much from a straight line trajectory towards their respective targets, and arrive at a dense configuration similar to that in Fig. \ref{fig:dubins_result}.}
  \label{fig:cc_traj}
\end{figure}

\subsubsection{Centralized Control}
Fig. \ref{fig:cc_traj} shows the simulated trajectories in the situation where a centralized controller enforces each vehicle to use the optimal controller $\ctrl^\text{dstb}_i(t, \state_i)$ according to \eqref{eq:opt_ctrl_i}, as described in Section \ref{sec:cc}. In this case, vehicles appear to deviate slightly from a straight line trajectory towards their respective targets, just enough to avoid higher-priority vehicles. The deviation is small since the centralized controller is quite restrictive, making the possible positions of higher-priority vehicles cover a small area. In the dense configuration at $t=-1.0$, the vehicles are close to each other but still outside each other's danger zones.

\begin{figure}[H]
  \centering
  \includegraphics[width=\columnwidth]{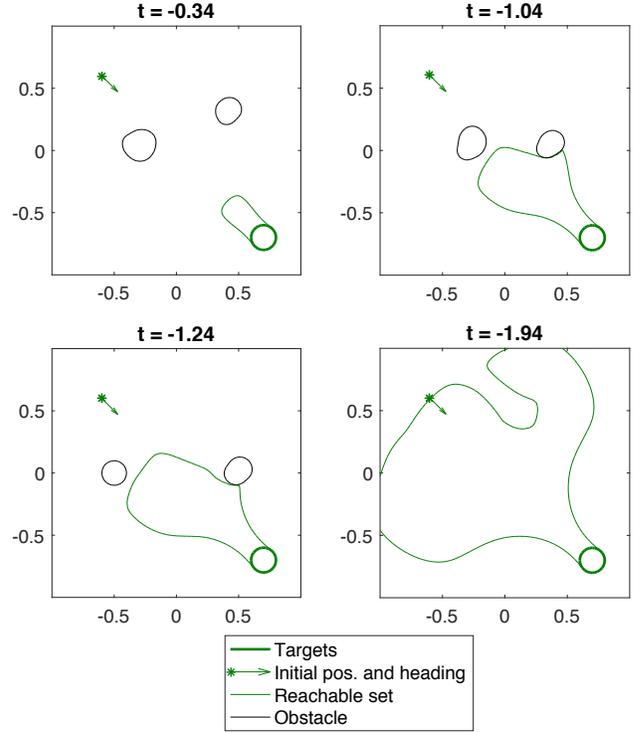}
  \caption{Evolution of the BRS and the obstacles induced by $\veh_1$ and $\veh_2$ for $\veh_3$ in the centralized control method. Since vehicles apply the optimal control at all times, the obstacle sizes are only slightly bigger than those in Fig. \ref{fig:dubins_reach_all} and \ref{fig:dubins_reach_3}.}
  \label{fig:cc_rs3}
\end{figure}

Fig. \ref{fig:cc_rs3} shows the evolution of the BRS for $\veh_3$ (green boundary), as well as the obstacles (black boundary) induced by the higher-priority vehicles $\veh_1$ and $\veh_2$. The locations of the induced obstacles at different time points include the actual positions of $\veh_1$ and $\veh_2$ at those times, and the sizes of obstacles remain relatively small. The $\ldt_i$ values for the four vehicles (in order) in this case are $-1.35, -1.37, -1.94$ and $-2.04$, relatively close for vehicles pairs $(\veh_1, \veh_2)$ and $(\veh_3, \veh_4)$, because the obstacles generated by higher-priority vehicles are small and hence do not affect the $\ldt_i$ of lower-priority vehicles significantly.
\subsubsection{Least Restrictive Control}
\begin{figure}
  \centering
  \includegraphics[width=\columnwidth]{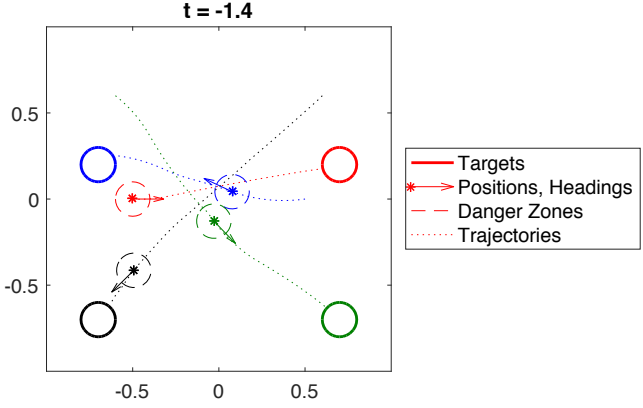}
  \caption{Simulated trajectories in the least restrictive control method. All vehicles start moving before $\veh_1$ starts, because the large obstacles make it optimal to wait until higher priority vehicles pass by, leading to earlier $\ldt_i$'s. }
  \label{fig:lrc_traj}
\end{figure}

\begin{figure}
  \centering
  \includegraphics[width=\columnwidth]{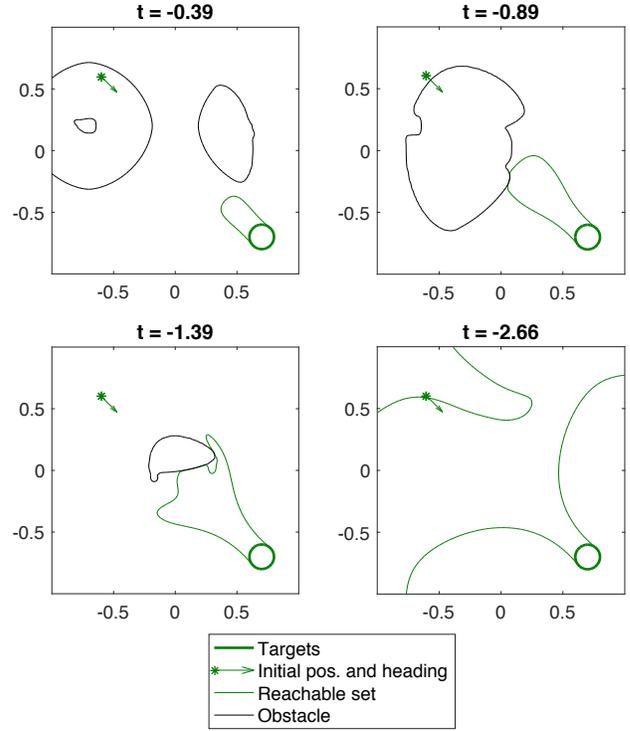}
  \caption{Evolution of the BRS for $\veh_3$ in the least restrictive control method. In this case, $\ldt_3=-2.66$, significantly earlier than that in the centralized control method ($-1.94$), reflecting the impact of larger induced obstacles.}
  \label{fig:lrc_rs3}
\end{figure}

Fig. \ref{fig:lrc_traj} shows the simulated trajectories in the situation where each vehicle assumes that higher-priority vehicles use the least restrictive control to reach their targets, as described in \ref{sec:lrc}. Fig. \ref{fig:lrc_rs3} shows the BRS and induced obstacles for $\veh_3$.

$\veh_1$ (red) takes a relatively straight path to reach its target. From the perspective of all other vehicles, large obstacles are induced by $\veh_1$, since lower-priority vehicles make the weak assumption that higher-priority vehicles are using the least restrictive control. Because the obstacles induced by higher-priority vehicles are so large, it is faster for lower-priority vehicles to wait until higher-priority vehicles pass by than to move around the higher-priority vehicles. As a result, the vehicles never form a dense configuration, and their trajectories are all relatively straight, indicating that they end up taking a short path to the target after higher-priority vehicles pass by. This is also indicated by the early $\ldt_i$ values for the four vehicles, $-1.35, -1.97, -2.66$ and $-3.39$, respectively. Compared to the centralized control method, $\ldt_i$'s are significantly earlier for all vehicles, except $\veh_1$, the highest-priority vehicle, since it need not account for any moving obstacles. 

From $\veh_3$'s (green) perspective, the large obstacles induced by $\veh_1$ and $\veh_2$ are shown in Fig. \ref{fig:lrc_rs3} as the black boundary. As the BRS (green boundary) evolves over time, its growth gets inhibited by the large obstacles for a long time, from $t=-0.89$ to $t=-1.39$. Eventually, the boundary of the BRS reaches the initial state of $\veh_3$ at $t = \ldt_3 = -2.66$.

\subsubsection{Robust Trajectory Tracking}
In the planning phase, we reduced the maximum turn rate of the vehicles from $1$ to $0.6$, and the speed range from $[0.5, 1]$ to exactly $0.75$ (constant speed). With these reduced control authorities, we determined from the disturbance rejection phase that a nominal trajectory from the planning phase can be robustly tracked within a distance of $R_{\text{EB}} = 0.075$.

\begin{figure}
  \centering
  \includegraphics[width=\columnwidth]{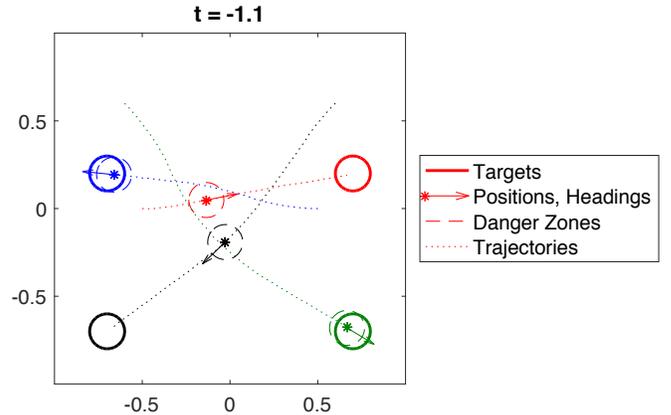}
  \caption{Simulated trajectories for the robust trajectory tracking method.}
  \label{fig:rtt_traj}
\end{figure}

\begin{figure}
  \centering
  \includegraphics[width=\columnwidth]{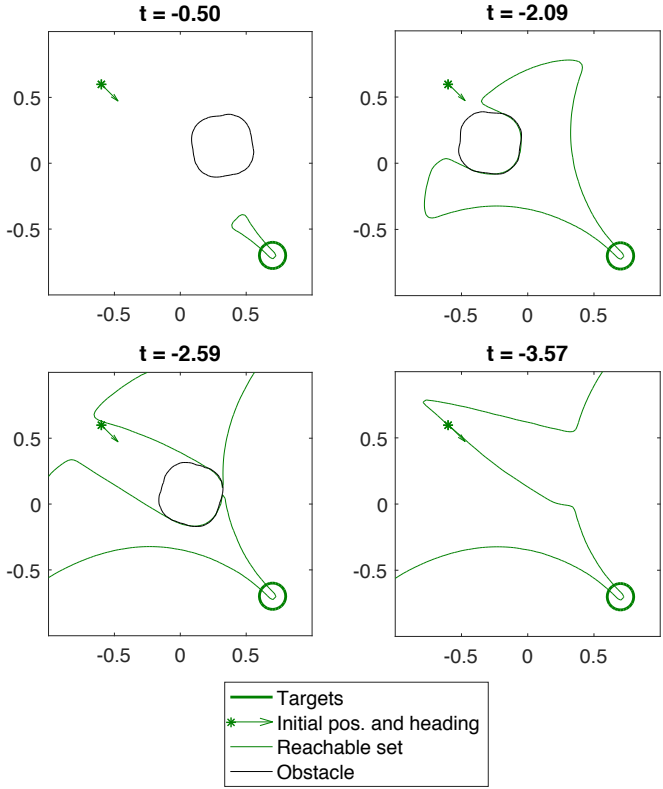}
  \caption{Evolution of the BRS for $\veh_3$ in the robust trajectory tracking method. As the BRS grows in time, the induced obstacles carve out a channel. Note that a smaller target set is used to compute the BRS to ensure that the vehicle reaches the target set by $t=0$ for any allowed tracking error.}
  \label{fig:rtt_rs3}
\end{figure}

Fig. \ref{fig:rtt_traj} shows the vehicle trajectories in the situation where each vehicle robustly tracks a pre-specified trajectory and is guaranteed to stay inside a ``bubble" around the trajectory. Fig. \ref{fig:rtt_rs3} shows the evolution of BRS and induced obstacles for vehicle $\veh_3$. The obstacles induced by other vehicles inhibit the evolution of the BRS, carving out thin “channels,” which can be seen at $t = -2.59$, that separate the BRS into different “islands”. 

In this case, the $\ldt_i$ values for the four vehicles are $-1.61, -3.16, -3.57$ and $-2.47$ respectively. In this method, vehicles use reduced control authority for path planning towards a reduced-size effective target set. As a result, higher-priority vehicles tend to have earlier $\ldt_i$'s compared to the other two methods, as evident from $\ldt_1$. Because of this ``sacrifice" made by the higher-priority vehicles during the path planning phase, the $\ldt$'s of lower-priority vehicles may be later compared to those in the other methods, as evident from $\ldt_4$. Overall, it is unclear how $\ldt_i$ will change for a vehicle compared to the other methods, as the conservative path planning leads to earlier $\ldt_i$'s for higher-priority vehicles and later $\ldt_i$'s for lower-priority vehicles.


\section{SPP With An Intruder \label{sec:intruder}}
In Section \ref{sec:incomp}, we made the basic SPP algorithm more robust by taking into account disturbances and considering situations in which vehicles may not have complete information about the control strategy of the other vehicles. However, if a vehicle not in the set of SPP vehicles enters the system, or even worse, if this vehicle is an adversarial intruder, the original plan can lead to vehicles entering into another vehicle's danger zone. If vehicles do not plan with an additional safety margin that takes a potential intruder into account, a vehicle trying to avoid the intruder may effectively become an intruder itself, leading to a domino effect. In this section, we propose a method to allow vehicles to avoid an intruder while maintaining the SPP structure.

\subsection{Theory}
In general, the effect of an intruder on the vehicles in structured flight can be entirely unpredictable, since the intruder in principle could be adversarial in nature, and the number of intruders could be arbitrary. Therefore, for our analysis to produce reasonable results, two assumptions about the intruders must be made.

\begin{assumption}
\label{as:avoidOnce}
At most one intruder (denoted as $\veh_I$) affects the SPP vehicles at any given time. The intruder is removed from the system after affecting the SPP vehicles after a duration of $\iat$. The removal of the intruder can be done, for example, by forcing it out of the altitude range of the SPP vehicles.
\end{assumption}

Let the time at which intruder appears in the system be $\tsa$ and the time at which it disappears be $\tea$. Assumption \ref{as:avoidOnce} implies that $\tea \leq \tsa + \iat$. Thus, any vehicle $\veh_i$ would need to avoid the intruder $\veh_{\intr}$ for a maximum duration of $\iat$. This assumption can be valid in situations where intruders are rare, and that some fail-safe or enforcement mechanism exists to force the intruder out of the altitude level affecting the SPP vehicles. Note that we do not make any assumptions about $\tsa$; however, we assume that once it appears, it stays for a maximum duration of $\iat$.

\begin{assumption}
\label{as:dynKnown}
The dynamics of the intruder are known and given by $\dot\state_\intr = f_\intr(\state_\intr, \ctrl_\intr, \dstb_\intr)$.
\end{assumption}

Assumption \ref{as:dynKnown} is required for HJ reachability analysis. In situations where the dynamics of the intruder are not known exactly, a conservative model of the intruder may be used instead.

Based on the above assumptions, we aim to design a control policy that ensures for each SPP vehicle separation with the intruder and with other SPP vehicles, and successful transit to the destination. However, depending on the initial state of the intruder, its control policy, and the disturbances in the dynamics of a vehicle and the intruder, a vehicle may arrive at different states after avoiding the intruder. Therefore, a control policy that ensures a successful transit to the destination needs to account for all such possible states, which is a path planning problem with multiple (infinite, to be precise) initial states and a single destination, and is hard to solve in general. 

Thus, we divide the intruder avoidance problem into two sub-problems. (i) We first design a control policy that ensures a successful transit to the destination if no intruder appears or successful avoidance of the intruder if it does appear. (ii) After the intruder is removed from the system at $\tea$, we solve a new SPP problem (that is, we ``re-plan") for vehicles which needed to avoid the intruder. In this case, the affected vehicles will re-plan as lowest-priority vehicles starting from the initial they happen to arrive at after avoiding the intruder. 

Suppose some vehicle $\veh_i$ starts avoiding the intruder $\veh_{\intr}$ at some time $t = \tsa$, and stops avoiding at $t = \tea$. When $t < \tsa$, $\veh_i$ must plan its path taking into account the possibility that it may need to avoid an intruder $\veh_\intr$. Since $\veh_i$ may spend a duration of up to $\iat$ performing avoidance, its induced obstacles $\ioset_k^i(t), k>i$ need to be computed in a way that reflects this possibility. The induced obstacles computation is discussed in Section \ref{sec:intruder_iocomp}.

We must also ensure that while avoiding the intruder, $\veh_i$ does not collide with the total obstacle set $\obsset_i(t)$. This requires computing the augmented total obstacle $\tilde\obsset_i(t)$; the computation of $\tilde\obsset_i(t)$ and the controller that guarantees the avoidance of the augmented obstacles are discussed in Section \ref{sec:intruder_aocomp}.

In Section \ref{sec:intruder_avoid}, we describe how $\veh_i$ can guarantee collision avoidance with the intruder. A pairwise collision avoidance problem such as this has been solved in isolation in \cite{Mitchell05}.

Finally, when $t > \tea$, $\veh_i$ has already successfully avoided the intruder, but depending on the state it happens to arrive at after avoiding the intruder, it may need to re-plan its trajectory to reach the target safely. The re-planning process is discussed in Section \ref{sec:re-plan_method1}.

\subsubsection{Induced Obstacle Computation} \label{sec:intruder_iocomp}
The goal of this section is to compute, for each lower-priority vehicle $\veh_i$, the time-varying obstacles induced by each higher-priority vehicle $\veh_j, j < i$, denoted by $\ioset_i^j(t)$. As before, the total obstacle set $\obsset_i(t)$ can then be obtained using \eqref{eq:obsseti}. To compute the obstacle that $\veh_i$ needs to avoid at time $t$, it is sufficient to consider the scenarios where $\tsa \in [t-\iat, t]$. This is because if $\tsa < t - \iat$, then the SPP vehicles would already be in the re-planning phase at time $t$ and hence cannot be in conflict. 

Depending on the information known to a lower-priority vehicle $\veh_i$ about $\veh_j$'s control strategy, we can use one of the three methods described in Section \ref{sec:incomp} to compute the ``base" obstacles $\boset_j(t)$; these are the obstacles that would have been induced by $\veh_j$ in the absence of an intruder. The base obstacles are respectively given by \eqref{eqn:ccObs_help2}, \eqref{eqn:lrcObs3} and \eqref{eqn:rttObs} for the centralized control, least restrictive control and robust trajectory tracking methods.

The induced obstacles, $\ioset_i^j(t)$, are then given by the states that $\veh_j$ can reach while avoiding the intruder, starting from some state in $\boset_j(\tsa), \tsa \in [t-\iat, t]$. These states can be obtained by computing an FRS from the base obstacles.
\begin{equation} \label{eq:FRS_intObs1}
\begin{aligned}
\frs_{j}^{\mathcal{O}}(t-\tau, t) = & \{y: \exists \ctrl_j(\cdot) \in \cfset_j, \exists \dstb_j(\cdot) \in \dfset_j, \\
& \state_j(\cdot) \text{ satisfies \eqref{eq:dyn}}, \state_j(t-\tau) \in \boset_j(t-\tau), \\
& \state_j(t) = y\}.
\end{aligned}
\end{equation}
$\frs_{j}^{\mathcal{O}}(t-\tau, t)$ represents the set of all possible states that $\veh_j$ can reach after a duration of $\tau$ starting from inside $\boset_j(t-\tau)$. This FRS can be obtained by solving the HJ VI in \eqref{eq:HJIVI_FRS} with the following Hamiltonian:
\begin{equation}
\ham_{j}^{\mathcal{O}}(\state_j, \costate) = \max_{\ctrl_j \in \cset_j} \max_{\dstb_j \in \dset_j} \costate \cdot f_j (\state_j, \ctrl_j, \dstb_j) \label{intobs2}.
\end{equation} 
Since $\tau \in [0, \iat]$, the induced obstacles can be obtained as:
\begin{equation} \label{eq:intObs}
\begin{aligned}
\ioset_i^j(t) & = \{\state_i: \exists y \in \pfrs_j(t), \|\pos_i - y\|_2 \le \rc \}\\
\pfrs_j(t) & = \{p_j: \exists \npos_j, (p_j, \npos_j) \in \bigcup_{\tau \in [0, \iat]} \frs_{j}^{\mathcal{O}}(t-\tau, t) \}
\end{aligned}
\end{equation}

Note that by the definition of base obstacles, $\boset_j(t+\tau_2) \subseteq \frs_{j}^{\text{BO}}(t+\tau_1, t+\tau_2) ~\forall t, \tau_2 > \tau_1$, where $\frs_{j}^{\text{BO}}(t+\tau_1, t+\tau_2)$ denotes the FRS of $\boset_j(t+\tau_1)$ computed for a duration of $\tau_2-\tau_1$. Therefore, we have that $\frs_{j}^{\mathcal{O}}(t-\tau, t) \subseteq \frs_{j}^{\mathcal{O}}(t-\iat, \iat) ~\forall \tau \in [0, \iat]$. Thus, $\pfrs_j(t)$ in \eqref{eq:intObs} can be equivalently written as
\begin{equation} \label{eq:intObs_help1}
\pfrs_j(t) = \{p_j: \exists \npos_j, (p_j, \npos_j) \in \frs_{j}^{\mathcal{O}}(t-\iat, t) \}.
\end{equation}

%
%

\subsubsection{Augmented Obstacle Computation} \label{sec:intruder_aocomp}
We next need to ensure that $\veh_i$ doesn't collide with the obstacles $\obsset_i(\cdot)$ computed in Section \ref{sec:intruder_iocomp} even when it is avoiding the intruder. In particular, we can compute a region around the obstacles $\obsset_i(\cdot)$ such that for all disturbances, $\veh_i$ can avoid colliding with obstacles for $\iat$ seconds regardless of its avoidance control, if $\veh_i$ starts outside this region. Augmenting $\obsset_i(\cdot)$ with this region gives us the augmented obstacles, $\tilde\obsset_i(\cdot)$, that can then be used during the path planning of $\veh_i$ to ensure collision avoidance with $\obsset_i(\cdot)$.  

Suppose that the intruder appears in the system at some time time $\tsa = t - \iat + \tau, \tau \in [0, \iat]$. In this case, we need to ensure that $\veh_i$ does not collide with the obstacle $\obsset_i(t + \tau)$ at time $t + \tau$, regardless of its control $\ctrl_i(s)$ and disturbance $\dstb_i(s)$ for the time interval $s \in [\tsa, t + \tau]$. It is, therefore, sufficient to avoid the $\tau$-horizon BRS of $\obsset_i(t + \tau)$ at time $t$. This argument applies for all $\tau \in [0, \iat]$. Mathematically,

\begin{equation} \label{eqn:inducedobs}
\tilde\obsset_i(t) = \bigcup_{\tau \in [0, \iat]} \brs^{\mathcal{G}}_{i}(t, t+\tau)
\end{equation}
where $\brs^{\mathcal{G}}_{i}(t, t+\tau)$ represents BRS of $\obsset_i(t+\tau)$ computed backwards for $\tau$ seconds. Formally, 
\begin{equation} \label{eqn:inducedobs_help1}
\begin{aligned}
\brs^{\mathcal{G}}_{i}(t, t+\tau) = & \{y: \exists \ctrl_i(\cdot) \in \cfset_i, \exists \dstb_i(\cdot) \in \dfset_i, \\
& \state_i(\cdot) \text{ satisfies \eqref{eq:dyn}}, \state_i(t) = y, \\
& \exists s \in [t, t+\tau], \state_i(s) \in \obsset_i(s)\}.
\end{aligned}
\end{equation}

The Hamiltonian $\ham^{\mathcal{G}}_{i}$ to compute $\brs^{\mathcal{G}}_{i}(\cdot)$ is given by:
\begin{equation} \label{eqn:BRS_obsham}
\ham^{\mathcal{G}}_{i}(\state_i, \costate) = \min_{\ctrl_i \in \cset_i} \min_{\dstb_i \in \dset_i} \costate \cdot f_i (\state_i, \ctrl_i, \dstb_i)
\end{equation}

\begin{remark}
Note that if we use the robust trajectory tracking method to compute the base obstacles, we would need to augment the obstacles in \eqref{eqn:inducedobs} by the error bound of $\veh_i$, $\disckernel_i$, as discussed in section \ref{sec:rtt}.
\end{remark}

Finally, we compute a BRS $\brs^{\text{AO}}_{i}(t, \sta_i)$ for path planning that contains the initial state of $\veh_i$ while avoiding these augmented obstacles:
\begin{equation} \label{eqn:intrBRS1}
\begin{aligned}
\brs^{\text{AO}}_{i}(t, \sta_i) = & \{y: \exists \ctrl_i(\cdot) \in \cfset_i, \forall \dstb_i(\cdot) \in \dfset_i, \\
& \state_i(\cdot) \text{ satisfies \eqref{eq:dyn}}, \forall s \in [t, \sta_i], \state_i(s) \notin \tilde\obsset_i(s), \\
& \exists s \in [t, \sta_i], \state_i(s) \in \targetset_i, \state_i(t) = y \}.
\end{aligned}
\end{equation}
The Hamiltonian $\ham^{\text{AO}}_{i}$ to compute BRS in \eqref{eqn:intrBRS1} is given by:
\begin{equation} \label{eqn:BRSham}
\ham^{\text{AO}}_{i}(\state_i, \costate) = \min_{\ctrl_i \in \cset_i} \max_{\dstb_i \in \dset_i} \costate \cdot f_i (\state_i, \ctrl_i, \dstb_i)
\end{equation}

Note that $\brs^{\text{AO}}_{i}(\cdot)$ ensures goal satisfaction for $\veh_i$ in the absence of intruder. The goal satisfaction controller is given by:
\begin{equation}
{\ctrl^{\text{AO}}_{i}}(t, \state_i) = \arg \min_{\ctrl_i \in \cset_i} \max_{\dstb_i \in \dset_i} \costate \cdot f_i (\state_i, \ctrl_i, \dstb_i)
\end{equation}
Moreover, if $\veh_i$ starts within $\brs^{\text{AO}}_{i}$, it is guaranteed to avoid collision for a duration of $\iat$, starting at any $\tsa < \sta_i$, irrespective of the control and disturbance applied during this time period. 

\subsubsection{Optimal Avoidance Controller} \label{sec:intruder_avoid}
First, we define relative dynamics of the intruder $\veh_{\intr}$ with state $\state_\intr$ with respect to $\veh_i$ with state $\state_i$.

\begin{equation}
\label{eq:reldyn}
\begin{aligned}
\state_{\intr, i} &= \state_\intr - \state_i \\
\dot \state_{\intr, i} &= f_r(\state_{\intr, i}, \ctrl_i, \ctrl_\intr, \dstb_i, \dstb_\intr)
\end{aligned}
\end{equation}

Given the relative dynamics, we compute the set of states from which the joint states of $\veh_{\intr}$ and $\veh_i$ can enter danger zone $\dz_{i\intr}$ despite the best efforts of $\veh_i$ to avoid $\veh_{\intr}$. This set of states is given by the BRS $\brs^{\text{CA}}(t, \iat),~ t \in [0, \iat]$:

\begin{equation} \label{eqn:optAvoid}
\begin{aligned}
\brs^{\text{CA}}_{i}(t, \iat) = & \{y: \forall \ctrl_i(\cdot) \in \cfset_i, \exists \ctrl_\intr(\cdot) \in \cfset_\intr, \exists \dstb_i(\cdot) \in \dfset_i, \\
& \exists \dstb_\intr(\cdot) \in \dfset_\intr, \state_{\intr, i}(\cdot) \text{ satisfies \eqref{eq:reldyn}},\\
& \exists s \in [t, \iat], \state_{\intr, i}(s) \in \targetset^{\text{CA}}_{i}, \state_{\intr, i}(t) = y\},
\end{aligned}
\end{equation}
where $\targetset^{\text{CA}}_{i} = \{\state_{\intr, i}: \|\pos_{\intr, i}\|_2 \le \rc\}$, and the Hamiltonian for computing this BRS is given by

\begin{equation*}
\begin{aligned}
&H^{\text{CA}}_{i}(\state_{\intr, i}, \costate) = \max_{\ctrl_i \in \cset_i} \Big( \\
&\qquad \qquad \qquad \min_{\ctrl_\intr \in \cset_\intr, \dstb_i \in \dset_i, \dstb_\intr \in \dset_\intr} \costate \cdot f_r(\state_{\intr, i}, \ctrl_i, \ctrl_\intr, \dstb_i, \dstb_\intr) \Big)
\end{aligned}
\end{equation*}

Once the value function $\valfunc^{\text{CA}}_{i}(t, \state_{\intr, i})$ corresponding to the BRS $\brs^{\text{CA}}_{i}(t, \iat)$ is computed, the optimal avoidance control ${\ctrl^{\text{CA}}_{i}}$ can be obtained as:
\begin{equation} \label{eqn:optAvoidCtrl}
\begin{aligned}
&{\ctrl^{\text{CA}}_{i}}(t, \state_i, \state_\intr)  = \arg \max_{\ctrl_i \in \cset_i} \Big( \\
&\qquad \qquad \qquad \min_{\ctrl_\intr \in \cset_\intr, \dstb_i \in \dset_i, \dstb_\intr \in \dset_\intr} \costate \cdot f_r(\state_{\intr, i}, \ctrl_i, \ctrl_\intr, \dstb_i, \dstb_\intr) \Big)
\end{aligned}
\end{equation}

Under normal circumstances when the intruder $\veh_{\intr}$ is far away, we have $\valfunc^{\text{CA}}_{i}(0, \state_{\intr, i}) > 0$; as $\veh_{\intr}$ gets closer to $\veh_i$, $\valfunc^{\text{CA}}_{i}(0, \state_{\intr, i})$ decreases. If $\veh_i$ applies the control ${\ctrl^{\text{CA}}_{i}}$ when $\valfunc^{\text{CA}}_{i}(0, \state_{\intr, i}) = 0$, then collision avoidance between $\veh_i$ and $\veh_{\intr}$ is guaranteed for a duration of $\iat$ under the worst-case intruder control strategy.

In addition, obstacle augmentation \eqref{eqn:inducedobs} ensures that $\veh_i$ does not collide with $\obsset_i(\cdot)$ during the avoidance maneuver. 
The overall control policy for avoiding the intruder and collision with other vehicles is thus given by:
\begin{equation*}
{\ctrl^{\text{A}}_{i}}(t) = 
\left \{ 
\begin{array}{ll}
{\ctrl^{\text{AO}}_{i}}(t, \state_i) & t \leq \tsa\\
{\ctrl^{\text{CA}}_{i}}(t, \state_i, \state_\intr) & \tsa\leq t \leq \tea
\end{array}
\right.
\end{equation*}

\subsubsection{Replanning after intruder avoidance\label{sec:re-plan_method1}} 
After the intruder disappears, goal satisfaction controllers which ensure that the vehicles reach their destinations can be obtained by solving an SPP problem as described in Section \ref{sec:incomp}, where the starting states of the vehicles are now given by the states they end up in, denoted $\tilde{\state}_j^0$, after avoiding the intruder. Let the optimal control policy corresponding to this goal satisfaction controller be denoted ${\ctrl^{\text{L}}_{i}}(t, \state_i)$. The overall control policy that ensures intruder avoidance, collision avoidance with other vehicles, and successful transition to the destination is given by:

\begin{equation*}
\ctrl_i^*(t) = 
\left \{ 
\begin{array}{ll}
{\ctrl^{\text{A}}_{i}}(t, \state_i) & t \leq \tea\\
{\ctrl^{\text{L}}_{i}}(t, \state_i) & t > \tea
\end{array}
\right.
\end{equation*}

Note that in order to re-plan using a SPP method, we need to determine feasible $\sta_i$ for all vehicles. This can be done by computing an FRS:
\begin{equation} \label{eq:re-planFRS}
\begin{aligned} 
\frs_i^{\text{RP}}(\tea, t) = & \{y \in \R^{n_i}: \exists \ctrl_i(\cdot) \in \cfset_i, \forall \dstb_i(\cdot) \in \dfset_i, \\
& \state_i(\cdot) \text{ satisfies \eqref{eq:dyn}}, \state_i(\tea) = \tilde{\state}_i^0, \\
& \state_i(t) = y, \forall s \in [\tea, t], \state_i(s) \notin \obsset_i^{\text{RP}}(s) \},
\end{aligned}
\end{equation}
\noindent where $\tilde{\state}_i^0$ represents the state of $\veh_i$ at $t = \tea$; $\obsset_i^{\text{RP}}(\cdot)$ takes into account the fact that $\veh_i$ needs to avoid higher-priority vehicles $\veh_j, j<i$ and is defined in an way analogous to \eqref{eq:obsseti}.

The FRS in \eqref{eq:re-planFRS} can be obtained by solving 

\begin{equation}
\begin{aligned}
\max \Big\{&D_t \valfuncfwd^{\text{RP}}(t, \state_i) + \ham_i^{\text{RP}}(t, \state_i, \nabla \valfuncfwd^{\text{RP}}(t, \state_i)), \\
&\qquad - \obsfunc^{\text{RP}}(t, \state_i) - \valfuncfwd^{\text{RP}}(t, \state_i) \Big\} = 0\\
&\valfuncfwd^{\text{RP}}(\tsa, \state_i) = \max\{\fc^{\text{RP}}(\state_i), -\obsfunc^{\text{RP}}(\tsa, \state_i)\} \\
&\ham_i^{\text{RP}}(\state_i, \costate) = \max_{\ctrl_i \in \cset_i} \min_{\dstb_i \in \dset_i} \costate \cdot f_i (\state_i, \ctrl_i, \dstb_i)
\end{aligned}
\end{equation} 

\noindent where $\valfuncfwd^{\text{RP}}, \obsfunc^{\text{RP}}, \fc^{\text{RP}}$ represent the FRS, obstacles during re-planning, and the initial state of $\veh_i$, respectively. The new $\sta$ of $\veh_j$ is now given by the earliest time at which $\frs_j^{\text{RP}}(\tea, t)$ intersects the target set $\targetset_j$, $\sta_j := \arg \inf_t \{ \frs_j^{\text{RP}}(\tea, t) \cap \targetset_j \neq \emptyset \}$. Intuitively, this means that there exists a control policy which will steer the vehicle to its destination by that time, despite the worst case disturbance it might experience.

\begin{remark}
Note that we only need to re-plan the trajectories of the vehicles that are affected by the intruder. In particular, if $\valfunc^{\text{CA}}(0, \state_{\intr, i}(t)) > 0$ during the entire duration $t \in [\tsa, \tea]$ for a vehicle, then the vehicle would need not to apply any avoidance control, and hence re-planning would not be required for this vehicle. 
\end{remark}

\begin{remark}
In general, an intruder can be present in the system for much longer than $\iat$, as long as it is not affecting the SPP vehicles. $\tsa$ thus really corresponds to the time an intruder starts affecting a SPP vehicle.
\end{remark}

\begin{remark}
Note that even though we have presented the analysis for one intruder, the proposed method can handle multiple intruders as long as only one intruder is present \textit{at any given time}. 
\end{remark}

We conclude this section with the overall SPP algorithm that takes into account an intruder that may appear for a duration of $\iat$: 
\begin{alg}
\label{alg:intruder}
\textbf{Intruder Avoidance algorithm (offline planning)}: Given initial conditions $\state_i^0$, vehicle dynamics \eqref{eq:dyn}, intruder dynamics in Assumption \ref{as:dynKnown}, target sets $\targetset_i$, and static obstacles $\soset_i, i = 1\ldots, \N$, for each $i$,
\begin{enumerate}
\item determine the total obstacle set $\obsset_i(t)$, given in \eqref{eq:obsseti}. In the case $i=1$, $\obsset_i(t) = \soset_i ~ \forall t$;
\item compute the augmented obstacle set $\tilde\obsset_i(t)$ given by \eqref{eqn:inducedobs}, where $\brs^{\mathcal{G}}_{i}(0, \tau)$ is given by \eqref{eqn:inducedobs_help1};
\item given $\tilde\obsset_i(t)$, compute the BRS $\brs^{\text{AO}}_{i}(t, \sta_i)$ defined in \eqref{eqn:intrBRS1};
\item the optimal control to avoid the intruder can be obtained by computing $\brs^{\text{CA}}_{i}(t, \iat)$ in \eqref{eqn:optAvoid} and using \eqref{eqn:optAvoidCtrl};
\item the induced obstacles $\ioset_k^i(t)$ for each $k>i$ can be computed using \eqref{eq:intObs}.
\end{enumerate}

\textbf{Intruder Avoidance algorithm (online re-planning)}: For each vehicle $i$ which performed avoidance in response to the intruder,
\begin{enumerate}
\item compute $\frs_i^{\text{RP}}(\tea, t)$ using $\eqref{eq:re-planFRS}$. The new $\sta_i$ for $\veh_i$ is given by $\arg \inf_t \{ \frs_j^{\text{RP}}(\tea, t) \cap \targetset_j \neq \emptyset \}$;
\item given $\sta_i$, $\tilde{\state}_i^0$, vehicle dynamics \eqref{eq:dyn}, target set $\targetset_i$, and static obstacles $\soset_i, i = 1\ldots, \N$, use any of the three SPP methods discussed in Section \ref{sec:incomp} for re-planning. 
\end{enumerate}
\end{alg}
\subsection{Numerical Simulations \label{sec:intruder_results}}
To illustrate that our SPP method is robust with respect to disturbances as well as a single intruder that is present for a duration of $\iat$, we use a five-vehicle example in which one of the five vehicles is an intruder. We assume that each vehicle has the dynamics given in \eqref{eq:dyn_i}. For this example, we chose the parameters $\underline{v} = 0.1, \bar{v} = 1, \bar\omega = 1$, and disturbance bounds $d_{r} = 0.1, \bar{d_{\theta}} = 0.2$, which correspond to a 10\% uncertainty in the dynamics. 

The vehicles' initial states, scheduled times of arrival, and target sets are the same as those described in Section \ref{sec:sim_dstb}, except that in this example, we have increased the target radius to $r=0.15$. For illustrate purposes, we have chosen to use the robust trajectory tracking method described in Section \ref{sec:rtt} for the base obstacles' computation, and hence each vehicle tracks a nominal trajectory.

\begin{figure}
  \centering
  \includegraphics[width=\columnwidth]{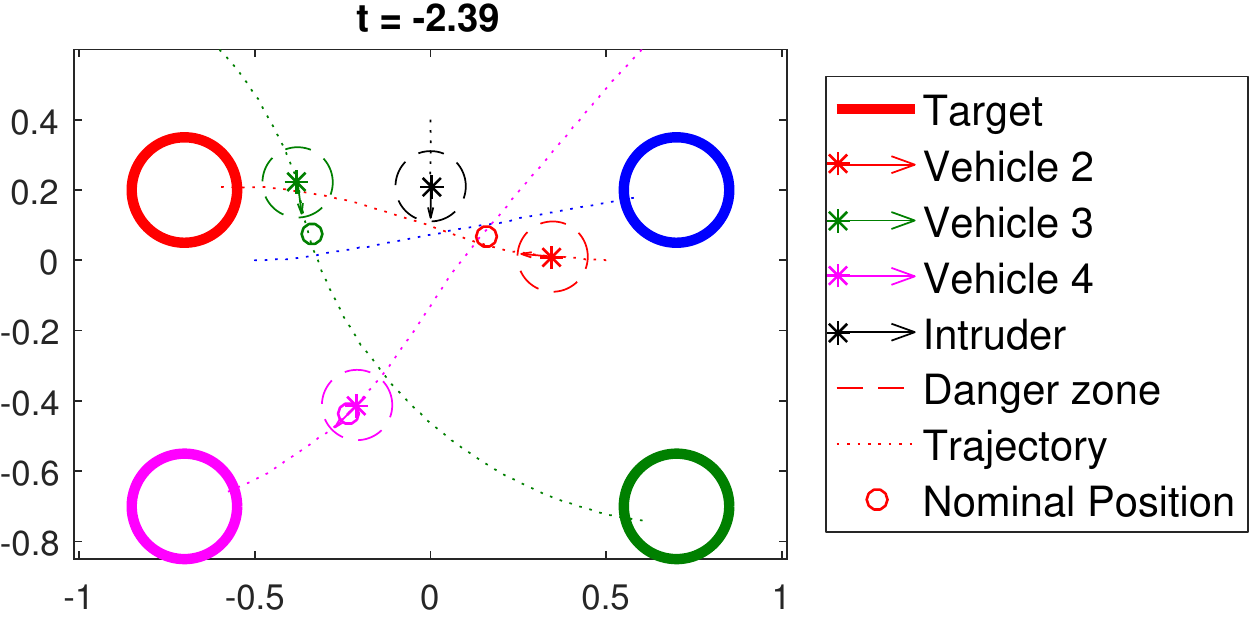}
  \caption{The positions of the SPP vehicles and the intruder at $t=-2.39$, the end of the intruder's appearance. The red and green vehicles $\veh_2, \veh_3$ have not been tracking their nominal trajectories for a while, and have been avoiding the intruder instead. Thus, their positions are far away from their nominal trajectories, indicated by the small colored circles. $\veh_4$ has not needed to avoid the intruder, and tracks its nominal trajectory closely. The nominal trajectory of $\veh_4$ allows it to stay far enough away from other vehicles so that all vehicles can remain safe in the presence of the intruder.}
  \label{fig:intruder1_traj}
\end{figure}

Fig. \ref{fig:intruder1_traj} shows the simulation at $t = \tea = -2.39$, which corresponds to the time at which the intruder ``disappears'' from the domain. This time is chosen to maximally highlight the impact of the intruder. Here, the intruder is shown in black, and the SPP vehicles are shown in the other different colors.

By the time $t = -2.39$, vehicle $\veh_2$ (red) and vehicle $\veh_3$ (green) have been avoiding the intruder for some time. This is evident from the amount of deviation between the actual positions of vehicles $\veh_2$ and $\veh_3$ (denoted by *) and their nominal positions (denoted by o) specified by the nominal trajectories they originally planned to track; these vehicles have abandoned nominal trajectory tracking in order to ensure safety with respect to the intruder. In contrast, $\veh_4$ (magenta), which has not needed to avoid the intruder, is tracking its nominal trajectory very closely (but not exactly, due to the presence of disturbances).

The SPP vehicles are rather far apart because a large margin is needed to ensure that they maintain separation even when multiple vehicles need to avoid the intruder. In this example in particular, the lowest-priority vehicle $\veh_4$ needed to depart very early compared to $\veh_2$ and $\veh_3$ so that if an intruder were to arrive, $\veh_4$ does not impede the ability of the other vehicles to perform avoidance. The early departure of $\veh_4$ can be inferred from the fact that at $t=-2.39$, it is already nearly at its target.

For the same reason, the highest-priority vehicle $\veh_1$ has not departed from its initial state yet, and thus is not shown at $t=-2.39$. $\veh_2$ and $\veh_3$ needed to depart very early compared to $\veh_1$ to ensure sufficient margin for avoidance maneuvers.

\begin{figure}
  \centering
  \includegraphics[width=\columnwidth]{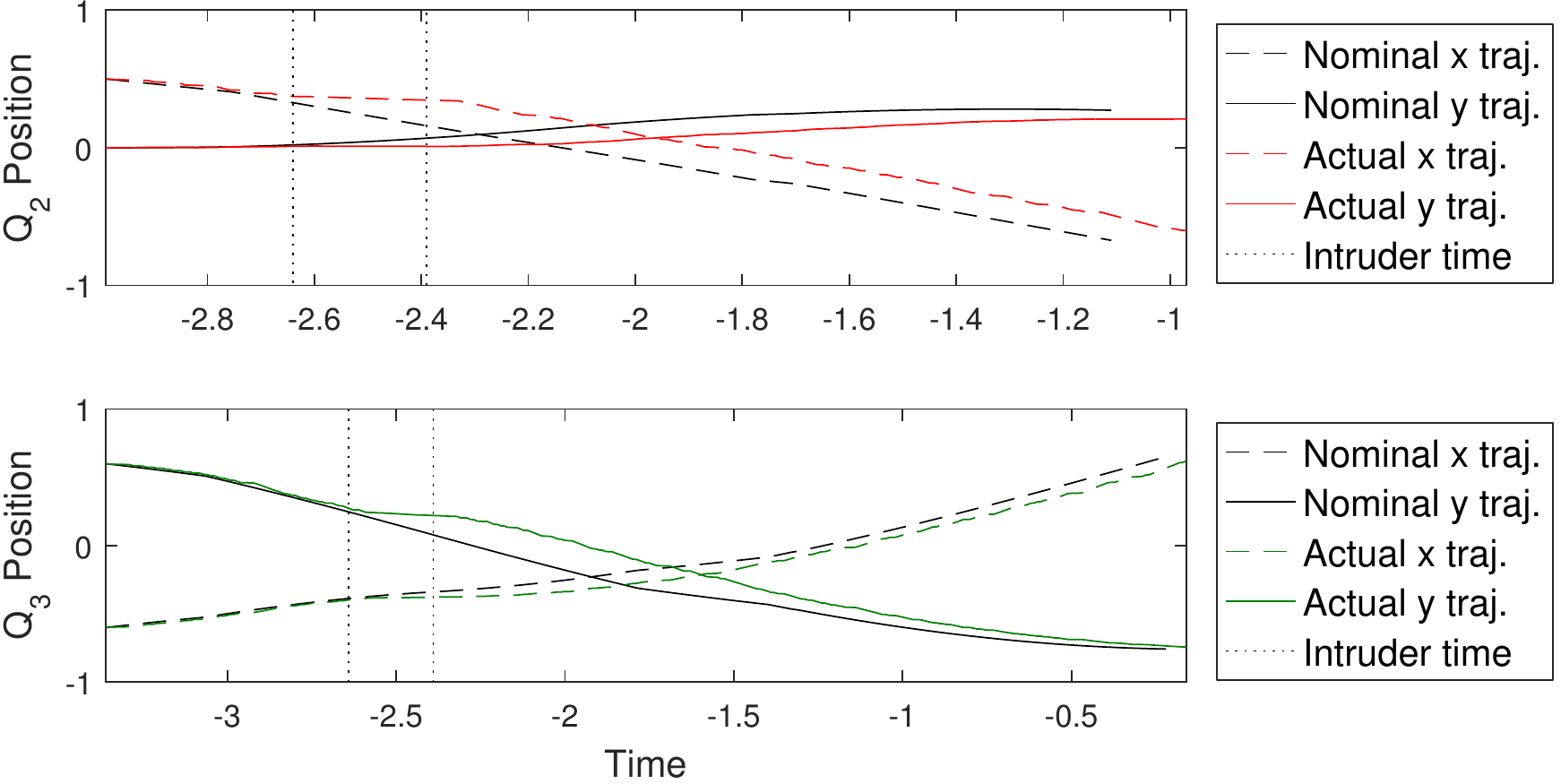}
  \caption{The difference between the initially planned nominal trajectories and the actual trajectories for vehicles $\veh_2$ (top subplot) and $\veh_3$ (bottom subplot), which needed to perform avoidance with respect to the intruder during the time interval marked by the vertical black dotted lines. Before the intruder's presence, both vehicles track their nominal trajectories closely; however, both vehicles later deviate significantly from their nominal trajectories in order to avoid the intruder. After the intruder is gone, both vehicles replan their trajectories and arrive at their targets at a later time.}
  \label{fig:intruder1_diff}
\end{figure}

Fig. \ref{fig:intruder1_diff} shows the nominal (black) and actual trajectories (red and green respectively) of vehicles $\veh_2$ (top subplot) and $\veh_3$ (bottom subplot). Specifically, the $x$ and $y$ positions over time are shown, and the black dotted vertical lines indicate the time interval in which the intruder is present. From Fig. \ref{fig:intruder1_diff}, one can clearly see that before the intruder was present, both vehicles are able to track their nominal trajectories closely. When the intruder appears, the vehicles deviate from their nominal trajectories significantly. After the intruder disappears, both vehicles re-plan new trajectories, and at a later time, the resulting actual trajectories eventually arrive at the same location as the nominal trajectories.

\section{Conclusions and Future Work}
Guaranteed-safe multi-vehicle path planning is a challenging problem, and previous analyses often either require strong assumptions on the motion of the vehicles or result in a large degree of conservatism. Optimal control and differential game techniques such as Hamilton-Jacobi (HJ) reachability are ideally suited for guaranteeing goal satisfaction and safety under disturbances, but become intractable for even a small number of vehicles.

Our robust sequential path planning (SPP) method assigns a strict priority ordering to vehicles to offer a tractable and practical approach to the multi-vehicle path planning problem. Under the proposed method, a portion of ``space-time'' is reserved for vehicles in the airspace in descending priority order to allow for dense vehicle configurations. Unlike previous priority-based methods, our approach accounts for disturbances and an adversarial intruder. SPP reduces the scaling of HJ reachability's computational complexity from exponential to linear with respect to the number of vehicles, while maintaining hard guarantees on goal satisfaction and safety under disturbances. In the presence of a single intruder vehicle, SPP still guarantees goal satisfaction and safety with a quadratically scaling computational complexity.

In the future, we plan to investigate ways of guaranteeing a maximum number of vehicles that need to re-plan, combine reachability analysis with other path planning methods to improve computation speed, and to better understand the scenarios under which the SPP scheme is the most useful by running large-scale simulations.

\bibliographystyle{IEEEtran}
\bibliography{IEEEabrv,references}
\end{document}